\documentclass[%
 reprint,
 amsmath,amssymb,
 aps,
]{revtex4-2}

\usepackage{graphicx}
\usepackage{dcolumn}
\usepackage{bm}

\usepackage{listings}



\begin{document}

\preprint{APS/123-QED}

\title{Fast proper orthogonal descriptors for many-body interatomic potentials}



\author{Ngoc-Cuong Nguyen}
\affiliation{%
Department of Aeronautics and Astronautics, Massachusetts Institute of Technology \\ 77 Massachusetts Avenue, Cambridge, MA 02139
}%


\date{\today}

\begin{abstract}
The development of differentiable invariant descriptors for accurate representations of atomic environments plays a central role in the success of interatomic potentials for chemistry and materials science. We introduce a method to generate fast proper orthogonal descriptors for the construction of many-body interatomic potentials and discuss its relation to exising empirical and machine learning interatomic potentials. A traditional way of implementing the proper orthogonal descriptors has a computational complexity that scales exponentially with the body order in terms of the number of neighbors. We present an algorithm to compute the proper orthogonal descriptors with a computational complexity that scales linearly with the number of neighbors irrespective of the body order. We show that our method can enable a more efficient implementation for a number of existing potentials and provide a scalable systematic framework to construct new many-body potentials. The new potentials are demonstrated on a data set of density functional theory calculations for Tantalum and compared with other interatomic potentials.  
\end{abstract}

\maketitle


\section{\label{sec:level1} Introduction}

Molecular dynamics (MD) simulations require an accurate  computation of energies and forces in order to analyze the physical movements of atoms. The two major approaches to computing interatomic energies and forces are (1) quantum mechanics (QM) calculations  and (2) interatomic potentials. While QM calculations are the most accurate modeling technique, they are restricted to analyzing small systems with thousands of atoms at most due to their high computational complexity. Interatomic potentials represent the potential energy surface (PES) of an atomic system as a function of atomic positions and thus leave out the detailed electronic structures. Since interatomic potentials typically have a computational complexity that scales linearly with the number of atoms, they can enable MD simulations of large systems with millions or even billions of atoms.


There is a long and successful history of using empirical interatomic potentials (EIPs) for MD simulations in chemistry and materials science. EIPs are usually derived from physical insights of electronic structure theories. More often than not, simple EIPs such as Lennard-Jones, Morse potentials do not accurately reproduce the results of QM calculations due to the use of simple, physically interpretable functional forms. Over the years, sophisticated many-body EIPs such as  the Finnis-Sinclair potential \cite{Finnis1984}, embedded atom method (EAM) \cite{Daw1984}, modified EAM (MEAM) \cite{Baskes1992}, Stillinger-Weber \cite{Stillinger1985}, Tersoff \cite{Tersoff1988}, EDIP \cite{Bazant1997}, REBO \cite{Brenner2002}, ReaxFF \cite{VanDuin2001} have been developed to treat a wide variety of atomic systems with different degrees of complexity. Despite many years of efforts devoted to the development of EIPs, there remain open problems to be addressed as regards to accuracy, efficiency, and transferability of EIPs. The development of EIPs is still an active and important area of research. 


\begin{figure}[htbp]
\includegraphics[scale=0.35]{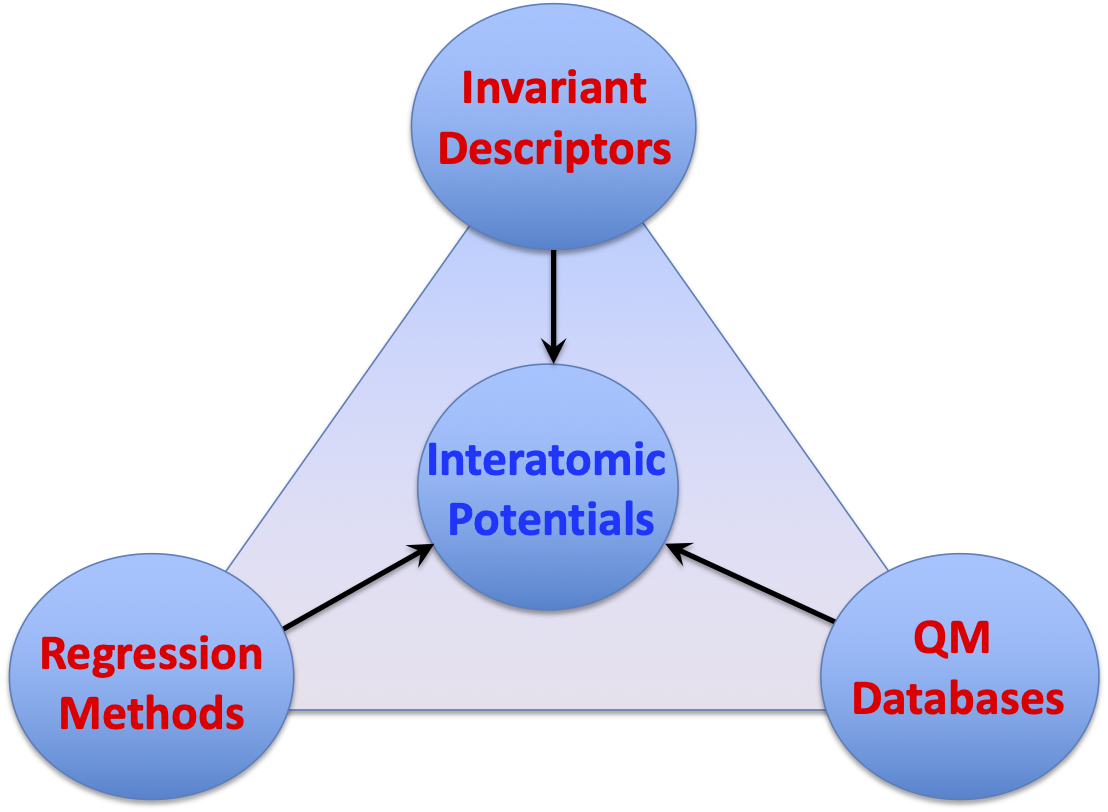}
\caption{\label{fig1} The main ingredients in the construction of machine learning interatomic potentials.}
\end{figure}

The past decade has seen a tremendous interest in machine learning interatomic potentials (MLIPs) due to their promising QM accuracy at significantly lower computational complexity than QM calculations \cite{Musil2021}. As illustrated in Figure \ref{fig1}, the main ingredients of any MLIPs include invariant descriptors, regression methods, and QM data sets. As the descriptors play a central role in the construction of accurate and efficient MLIPs, they must meet several requirements.  First and foremost, the descriptors must be invariant with respect to permutation, rotation, translation, and reflection of atoms in the system. Second, the descriptors must be differentiable with respect to the atomic positions to enable the calculation of  interatomic forces. Third, the descriptors must provide a detailed structural description of the local atomic environments to produce accurate and transferable MLIPs. Lastly, since the transformation from atom coordinates onto the descriptors has to be carried out for every atom, the computation of the descriptors has to be fast. 


In recent years, a wide variety of descriptors has been developed to represent atomic structures. As articulated in \cite{Musil2021}, there are two main approaches to mapping a configuration of atoms onto descriptors: atom density approach and internal coordinate approach. Examples of internal coordinate descriptors include permutation-invariant polynomials (PIPs) \cite{Braams2009,Nguyen2018a,VanDerOord2020}, atom-centered symmetry functions (ACSFs) \cite{Behler2007,Behler2011,Behler2014}, and proper orthogonal descriptors (PODs) \cite{Nguyen2022}. These internal coordinate descriptors are intrinsically invariant with respect to translation and rotation because they are functions of angles and distances. They are made to be permutationally invariant by summing symmetric functions over all possible atomic pairs and triplets within local atomic environments. However, achieving permutation invariance by such a way leads to the exponential scaling of the computational cost. The computational cost can be kept under control by restricting the range of interactions, the number of descriptors, and the body orders. 

A large number of existing descriptors follows the atom density approach to achieving invariant properties: rather than using internal coordinates that are inherently invariant to rotations and translations, they first describe a local atomic environment around a central atom  as an atom density function which is obtained by summing over localized functions centered on the relative positions of all atoms in the local environment. Such a density is naturally invariant to translation and permutation. The atomic neighborhood density is then expanded as a linear combination of appropriate basis functions, where the expansion coefficients are given by the inner products of the neighborhood density with the basis functions. The coefficients of the atomic neighborhood density are not directly useful as descriptors, because they are not invariant under rotation of the reference frame.  However, with an appropriate  basis set, one can construct rotationally invariant descriptors as appropriate sums of products of the density coefficients. In the atom density approach, the choices of the density function (e.g., the Dirac-delta function or Gaussian function) and the basis set (e.g., spherical harmonics combined with radial basis functions or hyperspherical harmonics) lead to different sets of descriptors. The power spectrum and bispectrum descriptors \cite{Bartok2013} use the Dirac-delta density function and spherical harmonics combined with radial basis functions, while the smooth overlap of atomic positions (SOAP) descriptors \cite{Bartok2013}  employ the Gaussian density function and the spectral neighbor analysis potential (SNAP) descriptors \cite{Thompson2015} are constructed from hyperspherical harmonics. The atomic cluster expansion (ACE) \cite{Drautz2019,Drautz2020} extends the power and bispectrum construction to obtain a complete set of invariant descriptors with arbitrary number of body orders. The  moment tensor potential (MTP) method \cite{Shapeev2016} projects the atomic density onto a tensor product of angular vectors to construct the moment tensors whose contraction results in invariant descriptors. The MTP descriptors  are related to the ACE descriptors through a change of basis \cite{Musil2021}. 

The main advantage of atom density descriptors is that their computational complexity scales linearly with the number of neighbors irrespective of the body orders.  This is to be contrasted with internal coordinate descriptors  whose computational complexity scales exponentially with the body order in terms of the number of neighbors. However, the cost of internal coordinate descriptors depends linearly on the number of basis functions, whereas that of atom density descriptors scales exponentially with the body order in terms of the number of basis functions.  This crucial difference makes atom density descriptors  more efficient than  internal coordinate descriptors when there are many neighbors and the body order is higher than 3. Despite this rather fundamental difference in construction and cost, the two families of descriptors can be shown essentially equivalent \cite{Musil2021}. Indeed, it is  possible to obtain internal coordinate descriptors using a contraction of density coefficients for a linear complete basis \cite{Musil2021}. It is important to point out that this equivalence is established only in the limit as the number of basis functions tends to infinity. For an incomplete basis, there will be residual difference between the two families of descriptors.

In this paper, we introduce a method to generate fast proper orthogonal descriptors (PODs). The fast PODs have flavors of both internal coordinate descriptors and atom density descriptors. In particular, the fast PODs are explicitly constructed from functions of internal coordinates, yet their computational complexity scales linearly with the number of neighbors irrespective of the body orders. Therefore, the fast PODs can be viewed either as internal coordinate descriptors or atom density descriptors, or both. As a result, the equivalence between the fast PODs and atom density descriptors can be shown even for a finite discrete basis.

The method introduced in this paper will have some significant implications for existing internal coordinate descriptors and empirical potentials. Most internal coordinate potentials and empirical potentials are three-body with a notable exception being the dihedral angle potentials that are four-body but involve selected groups of atoms rather than a sum over all possible triplets \cite{Musil2021}. This is because the cost of a naive implementation of these potentials scales quadratically (for three-body potentials) or cubically (for four-body potentials) with the number of neighbors. With the proposed method, it is possible to compute these existing potentials with a cost that scales  linearly with the number of neighbors. Our method will also open the possibility of constructing new internal coordinate and empirical potentials with  high body orders. In this paper, we present new internal coordinate potentials up to four body  and make some remarks about the construction of arbitrary-body potentials. 

The paper is organized as follows. In Section \ref{fastpod}, we introduce fast proper orthogonal descriptors and describe their implementation. In Section \ref{relationtoother}, we discuss the relationship among many empirical and machine learning interatomic potentials, and their connection to the present method. In Section \ref{regressionmethods}, we give an overview of several regression methods for constructing interatomic potentials from invariant descriptors. In Section \ref{results}, we present numerical results to demonstrate the newly developed potentials for Tantalum element. Finally, we provide some concluding remarks in Section \ref{conclusions}.

\section{Fast proper orthogonal descriptors}
\label{fastpod}

\subsection{Parametrized potential energy surface}

Let $\bm r_n \in \Omega$ be a position vector of an atom $n$ in a physical domain $\Omega \in \mathbb{R}^3$. We consider a system of $N$ atoms with $N$ position vectors $\bm R = (\bm r_1, \bm r_2, \ldots, \bm r_N) \in \mathbb{R}^{3N}$. The potential energy surface (PES) of the system of $N$ atoms can be expressed as a many-atom expansion of the form
\begin{equation}
\label{eq1}
\begin{split}
E(\bm R, \bm \mu)  =  & \sum_{i} V^{(1)}(\bm r_i, \bm \mu) + \frac12 \sum_{i,j} V^{(2)}(\bm r_i, \bm r_j, \bm \mu)  \\
& + \frac{1}{3!} \sum_{i,j,k} V^{(3)}(\bm r_i, \bm r_j, \bm r_k,  \bm \mu)  \\ 
& + \frac{1}{4!} \sum_{i,j,k,l} V^{(4)}(\bm r_i, \bm r_j, \bm r_k, \bm r_l,  \bm \mu) + \ldots 
\end{split}
\end{equation}
where $V^{(1)}$ is the one-body potential often used for representing external field or energy of isolated elements, and $V^{(2)}, V^{(3)}, V^{(4)}, \ldots$ are the higher-body potentials. The superscript on each potential denotes its body order. Note that each $q$-body potential $V^{(q)}$  depends on $\bm \mu$ which is a set of parameters to characterize the PES for a particular system. A separation of the PES into atomic contributions
yields
\begin{equation}
E(\bm R, \bm \mu) = \sum_{i=1}^N E_i(\bm R, \bm \mu)
\end{equation}
where $E_i$ is obtained from (\ref{eq1}) by removing the sum over index $i$. To make the PES invariant with respect to translation and rotation, the potentials should depend only on internal coordinates as follows 
\begin{equation}
\label{eq2}
\begin{split}
E_i  =  &  V^{(1)}(\bm \mu) + \frac12 \sum_{j} V^{(2)}( r_{ij}, \bm \mu) \ +  \\
& \frac{1}{3!} \sum_{j,k} V^{(3)}(r_{ij}, r_{ik}, w_{ijk}, \bm \mu) \ + \\ 
& \frac{1}{4!} \sum_{j,k,l} V^{(4)}(r_{ij}, r_{ik}, r_{ik}, w_{ijk}, w_{ijl}, w_{ikl}, \bm \mu) + \ldots  
\end{split}
\end{equation}
where $\bm r_{ij} = \bm r_j - \bm r_i$, $r_{ij} = |\bm r_{ij}|$, $w_{ijk} = \cos \theta_{ijk} = \hat{\bm r}_{ij} \cdot \hat{\bm r}_{ik}$, $\hat{\bm r} = \bm r/|\bm r|$. The internal coordinates include both distances $r_{ij}, r_{ik}, r_{il}$ and angles $w_{ijk}, w_{ijl}, w_{ikl}$. The number of internal coordinates for $V^{(q)}$ is equal to $(q-1)q/2$. In order to be invariant with respect to permutation, the potentials must be invariant with respect to permutation of the indices $j,k,l$.

The many-body expansion (\ref{eq2}) has rarely been used in the construction of interatomic potentials for $q \ge 4$. This is because of the exponential scaling of the cost to evaluate the sums in (\ref{eq2}). Even if a cut-off  is introduced in order to take into account only nearby atoms within a cutoff radius $r_{\rm cut}$, the evaluation of the sums associated with $V^{(q)}$ in (\ref{eq2}) scales as $N_i^{(q-1)}$, where  $N_i$ corresponds to a number of neighbors around atom $i$. Many problems require cut-off distances such that $N_i \ge  100$, which makes the cost of evaluating the sums in  (\ref{eq2}) too expensive for $q \ge 4$. One of the main contributions in this paper is to develop internal coordinate descriptors and associated many-body potentials that can be computed with linear complexity in terms of $N_i$.

\subsection{Two-body proper orthogonal descriptors}
\label{twobodypod}

Two-body PODs are exactly the same two-body descriptors introduced in our previous work \cite{Nguyen2022}. We briefly describe the construction of two-body PODs for completeness and refer to \cite{Nguyen2022} for further details. We assume that the direct interaction between two atoms vanishes smoothly when their distance is greater than the cutoff distance $r_{\rm cut}$. Furthermore, we assume that two atoms can not get closer than the inner cutoff distance $r_{\rm in}$ due to  Pauli exclusion  principle. Letting $r \in (r_{\rm in}, r_{\rm cut})$, we introduce the following parametrized radial functions
\begin{equation}
\label{eq3}
\phi(r, r_{\rm in}, r_{\rm cut}, \alpha, \beta)  = \frac{\sin (\alpha \pi x) }{r - r_{\rm in}}, \qquad  \varphi(r, \gamma)  = \frac{1}{r^\gamma} ,    
\end{equation}
where the scaled distance function $x$ is given by
\begin{equation}
x(r, r_{\rm in}, r_{\rm cut}, \beta) = \frac{e^{-\beta(r - r_{\rm in})/(r_{\rm cut} - r_{\rm in})} - 1}{e^{-\beta} - 1} .
\end{equation}
We introduce the following function as a convex combination of the two functions in (\ref{eq3})
\begin{equation}
\psi(r, \bm \mu)  = \kappa \phi(r, r_{\rm in}, r_{\rm cut}, \alpha, \beta) + (1- \kappa)  \varphi(r, \gamma) ,
\end{equation}
where $\mu_1 = r_{\rm in}, \mu_2 = r_{\rm cut}, \mu_3 = \alpha, \mu_4 = \beta, \mu_5 = \gamma$, and $\mu_6 = \kappa$. The two-body parametrized potential is defined as follows
\begin{equation}
\label{eq6}
V^{(2)}(r_{ij}, \bm \mu)  = f_{\rm c}(r_{ij}, \bm \eta) \psi(r_{ij}, \bm \mu)
\end{equation}
where the cut-off function $f_{\rm c}(r_{ij}, \bm \eta)$ is 
\begin{equation}
\label{eq8}
 f_{\rm c}(r, \bm \mu)  =  \exp \left(1 -\frac{1}{\sqrt{\left(1 - \frac{(r-r_{\rm in})^3}{(r_{\rm cut} - r_{\rm in})^3} \right)^2 + \epsilon}} \right)
\end{equation}
with $\epsilon = 10^{-6}$. This cut-off function ensures the smooth vanishing of the two-body potential and its derivative for $r_{ij} \ge r_{\rm cut}$.

Based on the two-body parametrized potential (\ref{eq6}), we construct the proper orthogonal descriptors as follows. Given $L$ parameter tuples $\bm \mu_\ell, 1 \le \ell \le L$, we introduce the following set of snapshots 
\begin{equation}
\label{eq11}
\Phi_\ell(r_{ij}) =  V^{(2)}(r_{ij}, \bm \mu_\ell),  \quad \ell = 1, \ldots, L .
\end{equation}
We next employ the proper orthogonal decomposition \cite{Nguyen2022} to generate an orthogonal basis set which is known to be optimal for representation of the snapshot family $\{\Phi_\ell\}_{\ell=1}^L$.  In particular, the  orthogonal radial basis functions are computed as follows
\begin{equation}
\label{eq12}
R_k(r_{ij}) = \sum_{\ell = 1}^L Q_{\ell k} \,  \Phi_\ell(r_{ij}), \qquad k = 1, \ldots, K_r^{(2)} , 
\end{equation}
where the number of radial basis functions $K_r^{(2)}$ is chosen less than $L$ and $Q_{\ell k}$ are a  matrix whose columns are eigenvectors of an eigenvalue problem obtained using the proper orthogonal decomposition. We refer to \cite{Nguyen2022} for the detailed discussion about the sampling of the parameter tuples $\{\bm \mu_\ell\}_{\ell=1}^L$ and the proper orthogonal decomposition. The two-body proper orthogonal descriptors at each atom $i$ are computed by summing the orthogonal basis functions over the neighbors of atom $i$ as follows
\begin{equation}
\label{eq12b}
\mathcal{D}^{(2)}_{ik}  = \sum_{j=1}^{N_i} R_k(r_{ij}), \quad 1 \le k \le K_r^{(2)}. 
\end{equation}
The orthogonal basis functions $R_k(r_{ij})$ are also used to construct three-body and four-body PODs as we discuss next. 

\subsection{Three-body proper orthogonal descriptors}

The three-body basis functions are defined as follows
\begin{equation}
\label{eq14}
U^{(3)}_{mn}(r_{ij}, r_{ik}, w_{ijk}) = R_m(r_{ij}) R_m(r_{ik}) \left( w_{ijk} \right)^{n},
\end{equation}
for $1 \le m \le K_r^{(3)}, 0 \le n \le P_a^{(3)}$, where the $R_m$ are the orthogonal radial basis functions described in the previous section. Here $K_r^{(3)}$ is the number of radial basis functions used to construct the three-body basis functions, and $P_a^{(3)}$ is the highest degree of three-body angular polynomials. In this work, the three-body PODs at each atom $i$ are obtained by summing (\ref{eq14}) over the neighbors $j$ and $k$ of atom $i$ as
\begin{equation}
\label{eq15}
\mathcal{D}^{(3)}_{imn}  = \sum_{j=1}^{N_i} \sum_{k=1}^{N_i} U^{(3)}_{mn}(r_{ij}, r_{ik}, w_{ijk}) .
\end{equation}
In the previous work \cite{Nguyen2022}, the three-body PODs are defined as 
\begin{equation}
\label{eq15b}
{D}^{(3)}_{imn}  = \sum_{j=1}^{N_i} \sum_{k > j}^{N_i} U^{(3)}_{mn}(r_{ij}, r_{ik}, w_{ijk}) .
\end{equation}
The sum over index $k$ is a subtle difference between (\ref{eq15}) and (\ref{eq15b}). In fact, the descriptors in (\ref{eq15}) are related to those in (\ref{eq15b}) by
\begin{equation}
\label{eq15c}
\mathcal{D}^{(3)}_{imn}  = 2 {D}^{(3)}_{imn} + \sum_{j=1}^{N_i} \left( R_m(r_{ij}) \right)^2 ,
\end{equation}
where the last term on the right side is a two-body term. This two-body term is quadratic and thus higher order than the two-body descriptors in (\ref{eq12b}). In other words, each descriptor in (\ref{eq15}) is a sum of a three-body descriptor and a quadratic two-body descriptor, whereas each descriptor in (\ref{eq15b}) is only three-body. These quadratic two-body terms bring some extra features into the three-body descriptors (\ref{eq15}).

While the above descriptors are naturally invariant with respect to translation and rotation, they are also invariant with respect to permutation because the  basis functions in (\ref{eq14}) are invariant with respect to permutation of $j$ and $k$.  At first glance, the cost of evaluating each descriptor in (\ref{eq15}) scales as $O(N_i^2)$. In Section \ref{implementation}, we will reformulate  (\ref{eq15}) into an atom density representation that is more computationally efficient to evaluate than (\ref{eq15}). 

\subsection{Four-body proper orthogonal descriptors}

Let $P_a^{(4)}$ be the highest degree of four-body angular polynomials. For any given $n \le P_a^{(4)}$, we introduce the following four-body angular functions
\begin{equation}
\label{eq16a}
f_p(w_{ijk}, w_{ijl}, w_{ikl}) = \left( w_{ijk} \right)^{a} \left( w_{ijl} \right)^{b} \left( w_{ikl} \right)^{c},
\end{equation}
and
\begin{equation}
\label{eq19a}
\begin{split}
g_p(w_{ijk}, w_{ijl}, w_{ikl}) = & \left( w_{ijk} \right)^{a} \left( w_{ijl} \right)^{b} \left( w_{ikl} \right)^{c} \ +  \\
&  \left( w_{ijk} \right)^{c} \left( w_{ijl} \right)^{a} \left( w_{ikl} \right)^{b} \ + \\   
&  \left( w_{ijk} \right)^{b} \left( w_{ijl} \right)^{c} \left( w_{ikl} \right)^{a} 
\end{split}
\end{equation}
where $a,b,c$ are integers such that $a + b + c = n$ and $a \ge b \ge c \ge 0$. The four-body angular functions $f_p$ and $g_p$ are listed in Table \ref{table1} and Table \ref{table2} for $n = 0, 1, \ldots, 4$, respectively. We note that all of the functions $g_p$ are invariant with respect to permutation of indices $j, k, l$, whereas many of the $f_p$ are not. 

\begin{table}[htbp]
\caption{\label{table1}%
Four-body angular functions $f_p$\footnote{Many of these functions are  not invariant to permutation of indices $j,k,l$.}
}
\begin{ruledtabular}
\begin{tabular}{ll}
$n$ &
$f_p$ \\
\colrule
0 & $f_1 = 1$   \\
1 & $f_2 = w_{ijk}$\\[0.5ex]
2 & $f_3 = w^2_{ijk}, \quad f_4 = w_{ijk} w_{ijl}$ \\[0.5ex]
3 & $f_5 = w^3_{ijk}, \quad f_6 = w^2_{ijk} w_{ijl}$, \quad $f_7 = w_{ijk} w_{ijl} w_{ikl}$ \\[0.5ex]
4 & $f_8 = w^4_{ijk}, \quad f_9 = w^3_{ijk} w_{ijl}$, \quad $f_{10} = w^2_{ijk} w^2_{ijl}$, \\[0.5ex]
& $f_{11} = w^2_{ijk} w_{ijl} w_{ikl}$ \\
\end{tabular}
\end{ruledtabular}
\end{table}

\begin{table}[htbp]
\caption{\label{table2}%
Permutationally invariant angular functions $g_p$.}
\begin{ruledtabular}
\begin{tabular}{ll}
$n$ &
$g_p$ \\
\colrule
0 & $g_1 = 1$   \\
1 & $g_2 = w_{ijk} + w_{ijl} + w_{ikl}$\\[0.5ex]
2 & $g_3 = w^2_{ijk} + w^2_{ijl} + w^2_{ikl}$ \\[0.5ex] & $g_4 = w_{ijk} w_{ijl} + w_{ijk} w_{ikl} + w_{ijl} w_{ikl}$ \\[0.5ex]
3 & $g_5 = w^3_{ijk} + w^3_{ijl} + w^3_{ikl}$ \\[0.5ex] 
  & $g_6 = w^2_{ijk} w_{ijl} + w_{ijk} w_{ikl}^2 + w_{ijl}^2 w_{ikl}$ \\[0.5ex] 
  & $g_7 = w_{ijk} w_{ijl} w_{ikl}$ \\[0.5ex]
4 & $g_8 = w^4_{ijk} + w^4_{ijl} + w^4_{ikl}$ \\[0.5ex] 
  & $g_9 = w^3_{ijk} w_{ijl} + w_{ijk} w_{ikl}^3 + w_{ijl}^3 w_{ikl}$ \\[0.5ex] 
  & $g_{10} = w^2_{ijk} w^2_{ijl} + w^2_{ijk} w^2_{ikl} + w^2_{ijl} w^2_{ikl}$ \\[0.5ex]   
  & $g_{11} = w^2_{ijk} w_{ijl} w_{ikl} + w_{ijk} w_{ijl}^2 w_{ikl} + w_{ijk} w_{ijl} w^2_{ikl} $ \\  
\end{tabular}
\end{ruledtabular}
\end{table}

The four-body basis functions are defined as follows
\begin{equation}
\label{eq17a}
U^{(4)}_{mp} = R_m(r_{ij}) R_m(r_{ik}) R_m(r_{il}) f_p(w_{ijk}, w_{ijl}, w_{ikl})
\end{equation}
for $1 \le m \le K_r^{(4)}, 0 \le p < K_a^{(4)}$. Here $K_r^{(4)}$ is the number of radial basis functions used to construct the four-body basis functions, while $K_a^{(4)}$ is the number of four-body angular basis functions and depends only on $P_a^{(4)}$. The four-body PODs at each atom $i$ are obtained by summing $U^{(4)}_{mp}$ over the neighbors $j$, $k$, $l$ of atom $i$ as
\begin{equation}
\label{eq18a}
\mathcal{D}^{(4)}_{imp}  = \sum_{ \{j,k,l\}=1}^{N_i} U^{(4)}_{mp}(r_{ij}, r_{ik}, r_{il}, w_{ijk}, w_{ijl}, w_{ikl}) .
\end{equation}
Although $U^{(4)}_{mp}$ are not invariant with respect to permutation, $\mathcal{D}^{(4)}_{imp}$ are invariant with respect to permutation, translation, and rotation due to the summation over all neighbors for indices $j, k, l$, and due to the use of distance and angle coordinates. We could replace $f_p$ with $g_p$ in (\ref{eq17a}) to obtain another set of descriptors. However, because the resulting descriptors only differ from the original descriptors (\ref{eq18a}) by constant factors, we would not gain anything but increase the computational cost.

To be accompanied with the three-body PODs in (\ref{eq15b}), we also introduce the following four-body PODs
\begin{equation}
\label{eq20a}
\begin{split}
{D}^{(4)}_{imp}  = \sum_{ j=1,k >j,l > k}^{N_i} & R_m(r_{ij}) R_m(r_{ik}) R_m(r_{il}) \\
& \times g_p(w_{ijk}, w_{ijl}, w_{ikl})    
\end{split}
\end{equation}
The exclusive sums over indices $k > j, l > k$ require the use of permutationally invariant angular functions $g_p$ to make ${D}^{(4)}_{imp}$ invariant with respect to permutation. It can be shown that each  descriptor in (\ref{eq18a}) is equal to the corresponding descriptor in (\ref{eq20a}) plus additional two-body and three-body terms. These  two-body  and two-body terms bring some extra features into the four-body descriptors (\ref{eq18a}). Specifically, if we assume that atom $i$ has exactly two neighbors, then  the descriptors ${D}^{(4)}_{imp}$ would be zero, whereas $\mathcal{D}^{(4)}_{imp}$ are not. In this case, $\mathcal{D}^{(4)}_{imp}$ reduces to three-body descriptors which are higher order than the three-body descriptors $\mathcal{D}^{(3)}_{imp}$ described earlier. 


From the standpoint of computational cost, if one naively performs the triple sums to evaluate both $\mathcal{D}^{(4)}_{imp}$ and ${D}^{(4)}_{imp}$, then the latter would be several times less expensive than the former. This is one of the reasons that most internal coordinate potentials  use the exclusive sums to reduce the computational cost. We will actually show that   $\mathcal{D}^{(4)}_{imp}$ are more significantly  faster to compute than ${D}^{(4)}_{imp}$. In the next section, we show how to compute $\mathcal{D}^{(4)}_{imp}$ with a cost that scales linearly with $N_i$. This linear scaling cost can be  orders of magnitude less than $O(N^3_i)$, which is the complexity of the straightforward  implementation by performing the triple sums (\ref{eq18a}).


\subsection{Atom density representation}
\label{implementation}


We introduce an algorithm to compute the fast PODs by reformulating them into more computationally efficient forms. We begin by noting from  (\ref{eq14}) and (\ref{eq15})  that the three-body descriptors are explicitly expressed as
\begin{equation}
\label{eq22a}
\begin{split}
\mathcal{D}^{(3)}_{imn} = & \ \sum_{j=1}^{N_i} \sum_{k=1}^{N_i}  R_m(r_{ij}) R_m(r_{ik}) \\
& \times \left( \hat{x}_{ij} \hat{x}_{ik} + \hat{y}_{ij} \hat{y}_{ik} + \hat{z}_{ij} \hat{z}_{ik} \right)^n  
\end{split}
\end{equation}
where $\bm r_{ij} = (x_{ij}, y_{ij}, z_{ij})$, $r_{ij} = |\bm r_{ij}|$, and $(\hat{x}_{ij}, \hat{y}_{ij}, \hat{z}_{ij}) = (x_{ij}/r_{ij}, y_{ij}/r_{ij}, z_{ij}/r_{ij})$. Furthermore, we note from (\ref{eq16a}), (\ref{eq17a}) and (\ref{eq18a}) that the four-body descriptors are explicitly expressed as
\begin{equation}
\label{eq23a}
\begin{split}
\mathcal{D}^{(4)}_{imp} = & \ \sum_{j=1}^{N_i} \sum_{k=1}^{N_i} \sum_{l=1}^{N_i}  R_m(r_{ij}) R_m(r_{ik}) R_m(r_{il}) \\
& \times \left( \hat{x}_{ij} \hat{x}_{ik} + \hat{y}_{ij} \hat{y}_{ik} + \hat{z}_{ij} \hat{z}_{ik} \right)^a \\
& \times \left( \hat{x}_{ij} \hat{x}_{il} + \hat{y}_{ij} \hat{y}_{il} + \hat{z}_{ij} \hat{z}_{il} \right)^b \\
& \times \left( \hat{x}_{il} \hat{x}_{ik} + \hat{y}_{il} \hat{y}_{ik} + \hat{z}_{il} \hat{z}_{ik} \right)^c .
\end{split}
\end{equation}
In order to efficiently evaluate (\ref{eq22a}) and (\ref{eq23a}), we express them as the sums of  the products of atomic basis functions that are defined as follows.


Next, we introduce a basis set of angular monomials 
\begin{equation}
\label{eq23b}
A_{n \ell}(\hat{\bm r}_{ij} ) = \hat{x}_{ij}^a \ \hat{y}_{ij}^b \ \hat{z}_{ij}^c
\end{equation}
such that $a + b + c = n$ for any given integer $n \in [0,P_a]$, where $P_a = \max(P_a^{(3)}, P_a^{(4)})$ is the highest angular degree. Table \ref{table3} shows the basis set of angular monomials for $P_a = 4$. Note that the index $\ell$ satisfies  $0 \le \ell \le (n+1)(n+2)/2-1$ and the total number of angular monomials is $(P_a+1)(P_a+2)(P_a+3)/6$.  We now define the atomic basis functions at atom $i$ as the sum over all neighbors of atom $i$ of the products of radial basis functions and angular monomials  
\begin{equation}
\label{24b}
B_{imn \ell }  = \sum_{j=1}^{N_i} R_m(r_{ij})  A_{n \ell}(\hat{\bm r}_{ij}) .
\end{equation}
These atomic basis functions are used to rewrite  the descriptors as follows. 

\begin{table}[htbp]
\caption{\label{table3}
The basis set of angular monomials for $P_a = 4$.
}
\begin{ruledtabular}
\begin{tabular}{ll}
$n$ &
$A_{n \ell}(\hat{\bm r}_{ij} )$ \\
\colrule
0 & $1$ \\
1 & $\hat{x}_{ij}$, \ $\hat{y}_{ij}$, \ $\hat{z}_{ij}$ \\[0.5ex]
2 & $\hat{x}^2_{ij}$, \ $\hat{y}^2_{ij}$, \ $\hat{z}^2_{ij}$, \ $\hat{x}_{ij} \hat{y}_{ij}$, \ $\hat{x}_{ij} \hat{z}_{ij}$, \ $\hat{y}_{ij} \hat{z}_{ij}$ \\[0.5ex]
3 & $\hat{x}^3_{ij}$, \ $\hat{y}^3_{ij}$, \ $\hat{z}^3_{ij}$, \ $\hat{x}^2_{ij} \hat{y}_{ij}$, \ $\hat{x}^2_{ij} \hat{z}_{ij}$, \ $\hat{y}^2_{ij} \hat{x}_{ij}$, \ $\hat{y}^2_{ij} \hat{z}_{ij}$,  \\[0.5ex] & $\hat{z}^2_{ij} \hat{x}_{ij}$, $\hat{z}^2_{ij} \hat{y}_{ij}$, $\hat{x}_{ij} \hat{y}_{ij} \hat{z}_{ij}$ \\[0.5ex]  
4 & $\hat{x}^4_{ij}$, \ $\hat{y}^4_{ij}$, \ $\hat{z}^4_{ij}$, \ $\hat{x}^3_{ij} \hat{y}_{ij}$, \ $\hat{x}^3_{ij} \hat{z}_{ij}$, \ $\hat{y}^3_{ij} \hat{x}_{ij}$, \ $\hat{y}^3_{ij} \hat{z}_{ij}$, $\hat{z}^3_{ij} \hat{x}_{ij}$  \\[0.5ex] 
& $\hat{z}^3_{ij} \hat{y}_{ij}$, $\hat{x}^2_{ij} \hat{y}^2_{ij}$, \ $\hat{x}^2_{ij} \hat{z}^2_{ij}$, \ $\hat{y}^2_{ij} \hat{z}^2_{ij}$, \ $\hat{x}^2_{ij} \hat{y}_{ij} \hat{z}_{ij}$ \ $\hat{x}_{ij} \hat{y}^2_{ij} \hat{z}_{ij}$ \\[0.5ex]  
& $\hat{x}_{ij} \hat{y}_{ij} \hat{z}^2_{ij}$ \\[0.5ex]  
\end{tabular}
\end{ruledtabular}
\end{table}

\begin{table}[htbp]
\caption{\label{table4}
The multinomial coefficients of the three-body descriptors $\mathcal{D}^{(3)}_{imn}$ for $P_a^{(3)} = 4$.
}
\begin{ruledtabular}
\begin{tabular}{ll}
$n$ &
$c_{n \ell}$ \\
\colrule
0 & $1$ \\
1 & $1$, \ $1$, \ $1$ \\[0.5ex]
2 & $1$, \ $1$, \ $1$, \ $2$, \ $2$, \ $2$   \\[0.5ex]
3 & $1$, \ $1$, \ $1$, \ $3$, \ $3$, \ $3$, \ $3$, \ $3$, \ $3$, \ $6$    \\[0.5ex]
4 & $1$, \ $1$, \ $1$, \ $4$, \ $4$, \ $4$, \ $4$, \ $4$, \ $4$, \ $6$, \ $6$, \ $6$, \ $12$, \ $12$, \ $12$    \\[0.5ex]
\end{tabular}
\end{ruledtabular}
\end{table}

First, the two-body descriptors are computed as 
\begin{equation}
\mathcal{D}^{(2)}_{im} =  B_{im00} .      
\end{equation}
Next, we expand the angular component of the three-body descriptors as
\begin{equation*}
\left( \hat{x}_{ij} \hat{x}_{ik} + \hat{y}_{ij} \hat{y}_{ik} + \hat{z}_{ij} \hat{z}_{ik} \right)^n  = \sum_{\ell=0}^{L(n)} c_{n \ell} A_{n\ell}(\hat{\bm r}_{ij}) A_{n\ell}(\hat{\bm r}_{ik})
\end{equation*}
where $L(n) = (n+1)(n+2)/2-1$ and $c_{n\ell}$ correspond to the multinomial coefficients of the expansion of $(x+y+z)^n$. Table \ref{table4} shows the multinomial coefficients for $n=0,1,2,3,4$. Hence, the three-body descriptors (\ref{eq22a}) can be rewritten as 
\begin{equation}
\label{eq26a}
\mathcal{D}^{(3)}_{imn} =  \sum_{\ell=0}^{L(n)} c_{n \ell} B^2_{imn \ell}  \ .  
\end{equation}
Although (\ref{eq22a}) and (\ref{eq26a}) are mathematically equivalent, (\ref{eq26a}) is considerably less expensive to evaluate  than (\ref{eq22a}) when the number of neighbors is large.  


The reformulation of the four-body descriptors is more involved. For cases $b = c = 0$, we can write the four-body descriptors as
\begin{equation}
\label{eq27a}
\begin{split}
\mathcal{D}^{(4)}_{imp} = & \left( \sum_{l=1}^{N_i}  R_m(r_{il}) \right) \Biggl( \sum_{j=1}^{N_i} \sum_{k=1}^{N_i}  R_m(r_{ij}) R_m(r_{ik})  \\
&  \times \left( \hat{x}_{ij} \hat{x}_{ik} + \hat{y}_{ij} \hat{y}_{ik} + \hat{z}_{ij} \hat{z}_{ik} \right)^a  \Biggr) \\ 
= & \mathcal{D}^{(2)}_{im} \times  \mathcal{D}^{(3)}_{ima} \ .
\end{split}
\end{equation}
For other cases, we note that
\begin{equation*}
\begin{split}
(\xi_1 & + \xi_2 + \xi_3)^a  (\eta_1 + \eta_2 + \eta_3)^b (\zeta_1 + \zeta_2 + \zeta_3)^c  = \\
& \sum_{q =0}^{L(a)} \sum_{s=0}^{L(b)} \sum_{t=0}^{L(c)} c_{a p} c_{b s} c_{c t} A_{a q}(\bm \xi)  A_{b s}(\bm \eta) A_{c t}(\bm \zeta)
\end{split}
\end{equation*}
where $A_{aq}$ are the monomials defined in (\ref{eq23b}). By considering $\xi_1 = \hat{x}_{ij} \hat{x}_{ik}, \xi_2 = \hat{y}_{ij} \hat{y}_{ik}, \xi_3 = \hat{z}_{ij} \hat{z}_{ik}$, $\eta_1 = \hat{x}_{ij} \hat{x}_{il}, \eta_2 = \hat{y}_{ij} \hat{y}_{il}, \eta_3 = \hat{z}_{ij} \hat{z}_{il}$, $\zeta_1 = \hat{x}_{ik} \hat{x}_{il}, \zeta_2 = \hat{y}_{ik} \hat{y}_{il}, \zeta_3 = \hat{z}_{ik} \hat{z}_{il}$, we obtain
\begin{equation*}
A_{a q}(\bm \xi)  A_{b s}(\bm \eta) A_{c t}(\bm \zeta) =  A_{a' q'}(\hat{\bm r}_{ij})  A_{b' s'}(\hat{\bm r}_{ik})  A_{c' t'}(\hat{\bm r}_{il})  
\end{equation*}
where $a' = a + b$, $b' = a + c$, $c' = b + c$, and the index $q'$ depends on $q$ and $s$, $s'$  on $q$ and $t$, $t'$ on $s$ and $t$. Thus, we rewrite (\ref{eq23a}) as
\begin{equation*}
\label{eq29a}
\begin{split}
\mathcal{D}^{(4)}_{imp} = &  \sum_{j=1}^{N_i} \sum_{k=1}^{N_i} \sum_{l=1}^{N_i}  R_m(r_{ij}) R_m(r_{ik}) R_m(r_{il}) \ \times \\
&   \sum_{q =0}^{L(a)} \sum_{s=0}^{L(b)} \sum_{t=0}^{L(c)} c_{a q} c_{b s} c_{c t}  A_{a' q'}(\hat{\bm r}_{ij} ) A_{b' s'}(\hat{\bm r}_{ik} ) A_{c' t'}(\hat{\bm r}_{il} ) 
\end{split}
\end{equation*}
which reduces to
\begin{equation}
\label{eq30a}
\begin{split}
\mathcal{D}^{(4)}_{imp} = &  \sum_{q =0}^{L(a)} \sum_{s=0}^{L(b)} \sum_{t=0}^{L(c)} c_{a q} c_{b s} c_{c t} B_{im a' q'} B_{im b' s'} B_{im c' t'} .
\end{split}
\end{equation}
For $b=c=0$ the four-body descriptors are evaluated as the products of two-body and three-body descriptors according to (\ref{eq27a}). For other cases, they are evaluated according to (\ref{eq30a}). It is possible to make the calculation of the four-body descriptors faster by implementing $c = 0$ separately.

In summary, the internal coordinate PODs are rewritten as the atom density descriptors in terms of the sums of the products of the atomic basis functions. As discussed later, this atom density representation of the internal coordinate PODs has a strong connection with other atom density descriptors such as ACE \cite{Drautz2019,Drautz2020}  and MTP \cite{Shapeev2016}. The proposed method thus has flavors of both the internal coordinate approach and the atom density density approach. It is straightforward to extend the method to arbitrary body orders. For instance, the five-body descriptors can be constructed from four distance coordinates $r_{ij}, r_{ik}, r_{il}, r_{im}$ and six angle coordinates $w_{ijk}, w_{ijl}, w_{ijm}, w_{ikl}, w_{ikm}, w_{ilm}$ in a similar way as the four-body descriptors (\ref{eq23a}). In general, the $q$-body descriptors can be constructed from $(q-1)$ distance coordinates and $(q-2)(q-1)/2$ angle coordinates. Therefore, the method can construct a complete set of descriptors for any body orders.

\subsection{Computational complexity}

We discuss the computational cost of evaluating the fast PODs. The operation count for computing the atomic basis functions in (\ref{24b}) is $O(N_i K_r (P_a+1) (P_a+2) (P_a+3)/6)$ per atom, where $K_r = \max(K_r^{(3)}, K_r^{(4)})$ is the number of radial basis functions and $P_a = \max(P_a^{(3)}, P_a^{(4)})$ is the highest possible angular degree. The operation count for computing the three-body descriptors in (\ref{eq26a}) is $O(K_r^{(3)} (P_a^{(3)}+1) (P_a^{(3)}+2) (P_a^{(3)}+3)/6)$ per atom. The operation count for computing the four-body descriptors in (\ref{eq30a}) is $O(K_r^{(4)} C_a^{(4)})$ per atom, where $C_a^{(4)}$ is listed in Table \ref{table5} for $P_a^{(4)} = 0,1,\ldots,10$. Table \ref{table5} also lists the number of angular monomials $N_a^{(4)} = (P_a^{(4)}+1) (P_a^{(4)}+2) (P_a^{(4)}+3)/6$, the number of angular basis functions $K_a^{(4)}$, and the rounding to the nearest integer of $C_a^{(4)}/N_a^{(4)}$. If $N_i \approx \lfloor C_a^{(4)}/N_a^{(4)} \rceil$ then the operation count for computing the atomic basis functions in (\ref{24b}) is equal to that for computing the four-body descriptors in (\ref{eq30a}). For $N_i > \lfloor C_a^{(4)}/N_a^{(4)} \rceil$ (which is likely for most applications), the operation count for computing the atomic basis functions is dominant.

\begin{table}[htbp]
\caption{\label{table5}%
Number of angular basis functions $K_a^{(4)}$, number of angular monomials $N_a^{(4)}$, angular complexity $C_a^{(4)}$, and the ratio $\lfloor C_a^{(4)}/N_a^{(4)} \rceil$ as a function of $P_a^{(4)}$.
}
\begin{ruledtabular}
\begin{tabular}{ccccc}
$P_a^{(4)}$ &
$K_a^{(4)}$ & 
$N_a^{(4)}$ & 
$C_a^{(4)}$  &
$\lfloor C_a^{(4)}/N_a^{(4)} \rceil$  \\
\colrule
0  &         1  &         1    &       1 & 1 \\
1  &          2 &          4   &        4 & 1 \\
2  &         4 &          10   &       19 & 2\\
3  &         7 &          20   &       74 & 4 \\
4  &         11 &         35   &      209 & 6 \\
5  &         16 &         56   &      533 & 10 \\
6  &         23 &         84   &      1345 & 16 \\
7  &        31 &         120   &     2860 & 24 \\
8  &        41 &         165   &     5836 & 35 \\
9  &        53 &          220  &     11626 & 53 \\
10 &        67 &          286 &      21304 & 75 \\
\end{tabular}
\end{ruledtabular}
\end{table}
 
\subsection{Derivatives of the fast POD descriptors}

It is necessary to calculate derivatives of the descriptors in order to fit interatomic potentials or compute forces. The derivatives of the three-body descriptors with respect to the relative positions are calculated as
\begin{equation}
\label{eq26c}
\frac{\partial \mathcal{D}^{(3)}_{imn}}{\partial \bm r_{ij}} = 2 \sum_{\ell=0}^{L(n)} c_{n \ell} B_{imn \ell} \frac{\partial B_{imn \ell}}{\partial \bm r_{ij}}  , 
\end{equation}
where
\begin{equation}
\label{24c}
\frac{\partial B_{imn \ell}}{\partial \bm r_{ij}}  =  \frac{\partial R_m(r_{ij})}{\partial \bm r_{ij}}  A_{n \ell}(\hat{\bm r}_{ij})+ R_m(r_{ij}) \frac{\partial A_{n \ell}(\hat{\bm r}_{ij})}{\partial \bm r_{ij}}   .
\end{equation}
As the derivatives of the two-body and four-body descriptors are calculated similarly, they are not described to save space. The computational cost of calculating the derivatives of the descriptors is mostly dominated by the cost of computing $\frac{\partial B_{imn \ell}}{\partial \bm r_{ij}}$ in (\ref{24c}), which is only 3 times more than the cost of computing $B_{imn \ell}$. Therefore, the cost of calculating the derivatives of the descriptors is $O(N_i K_r (P_a+1) (P_a+2) (P_a+3)/2)$ per atom. 


\section{Empirical and machine learning interatomic potentials}
\label{relationtoother}

In this section, we present an overview of empirical and machine learning potentials. We show how the fast POD descriptors can be gainfully used in a number of existing EIPs to improve their performance. We propose some modifications of existing EIPs by using the fast POD descriptors to replace or augment some of the terms. These modifications aim at making some existing EIPs more efficient and accurate. We also survey machine learning potentials with particular emphasis on methods for constructing invariant descriptors. We discuss the relationship among the surveyed methods and their connection to the POD method.


\subsection{Empirical interatomic potentials}
\label{Empirical}


EIPs have been an active area of research in molecular dynamics. Empirical potentials are representations of the potential energy surface of the form (\ref{eq1}) in which the potentials are derived from physical
insights into electronic structure theories and the parameters are fitted to individual systems. A large collection of EIPs have been developed and extended by adding new terms into existing EIPs or calibrating new parameters for different systems. Pair potentials have some inherent limitations due to the use of simple two-body terms. Over the years, sophisticated many-body EIPs have been developed to treat a wide variety of atomic systems with different degrees of complexity.  The Stillinger-Weber potential \cite{Stillinger1985} has a two-body and three-body terms of the  form
\begin{equation*}    
\begin{split} E_i & =  \sum_{j > i} V^{(2)} (r_{ij}, \bm \mu) +
        \sum_{j \neq i} \sum_{k > j}
        V^{(3)} (r_{ij}, r_{ik}, \theta_{ijk}, \bm \mu) \\
V^{(2)}  & =  A_{ij} \epsilon_{ij} \left[ B_{ij} (\frac{\sigma_{ij}}{r_{ij}})^{p_{ij}} -
                  (\frac{\sigma_{ij}}{r_{ij}})^{q_{ij}} \right]
                  \exp \left( \frac{\sigma_{ij}}{r_{ij} - a_{ij} \sigma_{ij}} \right) \\
V^{(3)} & = \lambda_{ijk} \epsilon_{ijk} \left[ \cos \theta_{ijk} -
                  \cos \theta_{0ijk} \right]^2
                  \exp \left( \frac{\gamma_{ij} \sigma_{ij}}{r_{ij} - a_{ij} \sigma_{ij}} \right) \\
&                \quad    
                  \exp \left( \frac{\gamma_{ik} \sigma_{ik}}{r_{ik} - a_{ik} \sigma_{ik}} \right)
\end{split}
\end{equation*}
It was originally developed for pure Si, but has been extended to many other elements and compounds. EAM potentials \cite{Daw1984}
are commonly used to model metals with the following functional form 
\begin{equation}
E_i = F \left(\sum_{j \neq i}\ \rho (r_{ij}, \bm \mu)\right) +
      \frac{1}{2} \sum_{j \neq i} V^{(2)} (r_{ij}, \bm \mu)    
\end{equation}
where $F$ is the embedding function of the atomic electron density $\rho$. In fact,  EAM potentials have its root from the Finnis-Sinclair potential \cite{Finnis1984} in which the embedding function is a square root function. The multi-body nature of  EAM potentials is a result of the nonlinear embedding function.  Covalently bonded materials are often described by bond order potentials which have  a form that resembles a pair potential:
\begin{equation}
E_i  = \frac{1}{2} \sum_{j \neq i} \left[ V^{ (2)}_{\rm R}(r_{ij}, \bm \mu) + b_{ij}(\bm \mu) V^{ (2)}_{\rm A}(r_{ij}, \bm \mu) \right]   
\end{equation}
where $V^{ (2)}_{\rm R}$ and $V^{ (2)}_{\rm A}$ are the repulsive and attractive two-body terms, respectively; and $b_{ij}$ is a three-body term that depends on $r_{ij}, r_{ik}, \theta_{ijk}$. One of the most well-known bond order potentials is the Tersoff potential \cite{Tersoff1988} which has the following terms
\begin{equation}
\begin{split}
V^{ (2)}_{\rm R} & = f_{\rm c}(r_{ij})   A \exp (-\lambda_1 r_{ij}) \\
V^{ (2)}_{\rm A} & =  - f_{\rm c}(r_{ij})  B \exp (-\lambda_2 r_{ij}) \\
b_{ij} & =  \left( 1 + \beta^n {\zeta_{ij}}^n \right)^{-\frac{1}{2n}} \\
\zeta_{ij} & =  \sum_{k } f_{\rm c}(r_{ik}) g \left( \theta_{ijk}\right)
                 \exp \left[ {\lambda_3}^m (r_{ij} - r_{ik})^m \right] \\
g(\theta) & =  \gamma \left( 1 + \frac{c^2}{d^2} -
                \frac{c^2}{\left[ d^2 + (\cos \theta - \cos \theta_0)^2\right]} \right) .
\end{split}
\end{equation}
Since then several other EIPs have been developed by extending the formalism of the earlier work discussed above. Notably among them are MEAM \cite{Baskes1992}, EDIP \cite{Bazant1997}, REBO \cite{Brenner2002}, ReaxFF \cite{VanDuin2001}. 

The SW potential can be seen as a combination of the two-body and three-body POD descriptors with a particular choice of the radial and angular basis functions. If the embedding function  is a polynomial of degree $m$, $F(u) = \alpha_0 + \alpha_1 u^1 + \ldots + \alpha_m u^m$, then the body order of EAM potentials is $m+1$. In this case, the embedding term can be expressed as $\alpha_0 + \alpha_1 B_{i} + \ldots + \alpha_m B_i^m$, where $B_i = \sum_{j}\ \rho (r_{ij}, \bm \mu)$. This is a linear combination of the POD descriptors with $n = 0$, allowing the EAM potential to be represented by the POD descriptors. The MEAM potential \cite{Baskes1992}  was developed as a generalization of the EAM potential by including angular dependent interactions in the electron density term as follows
\begin{equation}
\rho_i = \sum_j \rho(r_{ij})
+ \sum_{jk} f(r_{ij}) f(r_{ik}) g\left(\cos(\theta_{ijk})\right)  ,  
\end{equation}
where $g$ is a polynomial. As a result, MEAM  can describe directional bonding, which is most apparent in covalent materials such as silicon and diamond. If the embedding function  is a polynomial of degree $m$, then the body order of MEAM potentials is $2m+1$. For instance, if $m = 2$ we have
\begin{equation}
\begin{split}
F(\rho_i) = & \Biggl(\sum_j \rho(r_{ij}) \Biggr)^2  
\\ & + 2 
 \sum_j \rho(r_{ij}) \sum_{jk} f(r_{ij}) f(r_{ik}) g\left(\cos(\theta_{ijk})\right)
 \\ & + \Biggl( \sum_{jk} f(r_{ij}) f(r_{ik}) g\left(\cos(\theta_{ijk})\right) \Biggr)^2.  
 \end{split}
\end{equation}
The first term is three-body, the second is four-body, and the last term is five-body. Indeed, the first term is a square of a two-body POD descriptor; the second term is the product of two-body and three-body POD descriptors, similar to (\ref{eq27a}); and the last term is nothing but a square of a three-body POD descriptor. As a result, MEAM can be expressed as a sum of products of two-body and three-body POD descriptors for any polynomial embedding function.

Traditionally, the SW and MEAM potentials are implemented by summing over neighbors of atom $i$ for both $j$ and $k$. This implementation results in $O(N_i^2)$ cost per atom. By expressing these potentials in terms of the fast POD descriptors, they can be implemented efficiently with the complexity of $O(N_i)$ only. However, bond order potentials such as the Tersoff potential can not be cast in terms of the fast POD descriptors. This is because the three-body term  of the Tersoff potential has the form
\begin{equation}
b_{ij} =  f \left( \sum_k  g(r_{ij}, r_{ik}, \theta_{ijk} ) \right)  
\end{equation}
where $f$ and $g$ are nonlinear functions. As such potential does not admit the atom density representation, its evaluation costs $O(N_i^2)$ per atom. To make a new bond order potential which can be computed with $O(N_i)$ per atom, we propose the following three-body term 
\begin{equation}
b_{ij} =  \sum_k \sum_{m} \sum_{n}   \alpha_{mn}f_m(r_{ij}) f_m( r_{ik}) \cos^n( \theta_{ijk} ) , 
\end{equation}
where $f_m(r_{ij}) = f_{\rm c}(r_{ij}) \exp \left( -{\beta}_m r_{ij}^m \right)$. Here $\alpha_{mn}$ and $\beta_m$ are parameters to be fitted.  

Most EIPs are  functions of two-body and three-body terms with a notable exception being the dihedral angle potentials that are four-body but involve selected groups of atoms rather than a sum over all possible triplets \cite{Musil2021}.  To construct efficient many-body potentials beyond three-body formalism, we make use of the POD method. For instance, the POD formalism can be used to define a new MEAM potential with the following form of the electron density
\begin{equation}
\begin{split}
\rho_i = & \sum_j \rho(r_{ij})
+ \sum_{jk} f(r_{ij}) f(r_{ik}) g\left(w_{ijk}\right) \\    
& + \sum_{jkl} d(r_{ij}) d(r_{ik}) d(r_{il}) h\left(w_{ijk}, w_{ijl}, w_{ikl}\right)
\end{split}
\end{equation}
where $\rho, f, d$ are any given functions and $g, h$ are polynomials. The resulting MEAM potential is $(3m+1)$-body, if the embedding function is a polynomial of degree $m$. 

With the POD method, it is possible to implement a number of many-body empirical potentials with a computational cost that scales only linearly with the number of neighbors. Furthermore, the method also brings about the possibility of constructing new EIPs, while maintaining linear scaling of the computational cost. One common trait of all EIPs is that the  functional forms and the number of parameters are always fixed, while the parameters are fitted to individual systems. This trait has pros and cons. On the one hand, EIPs are fast to evaluate energies and forces because they have a very small number of functional forms. On the other hand, one can not enlarge EIP models to improve their accuracy because the number of parameters and the functional forms are being fixed. EIPs are found to be significantly less accurate than MLIPs \cite{Thompson2015, Zuo2020}, mainly because MLIPs have the ability to adapt and enlarge the models to improve their predictability. However, MLIPs are less interpretable and generally less robust than EIPs.


\subsection{Machine learning interatomic potentials}

\begin{figure*}
\includegraphics[scale=0.5]{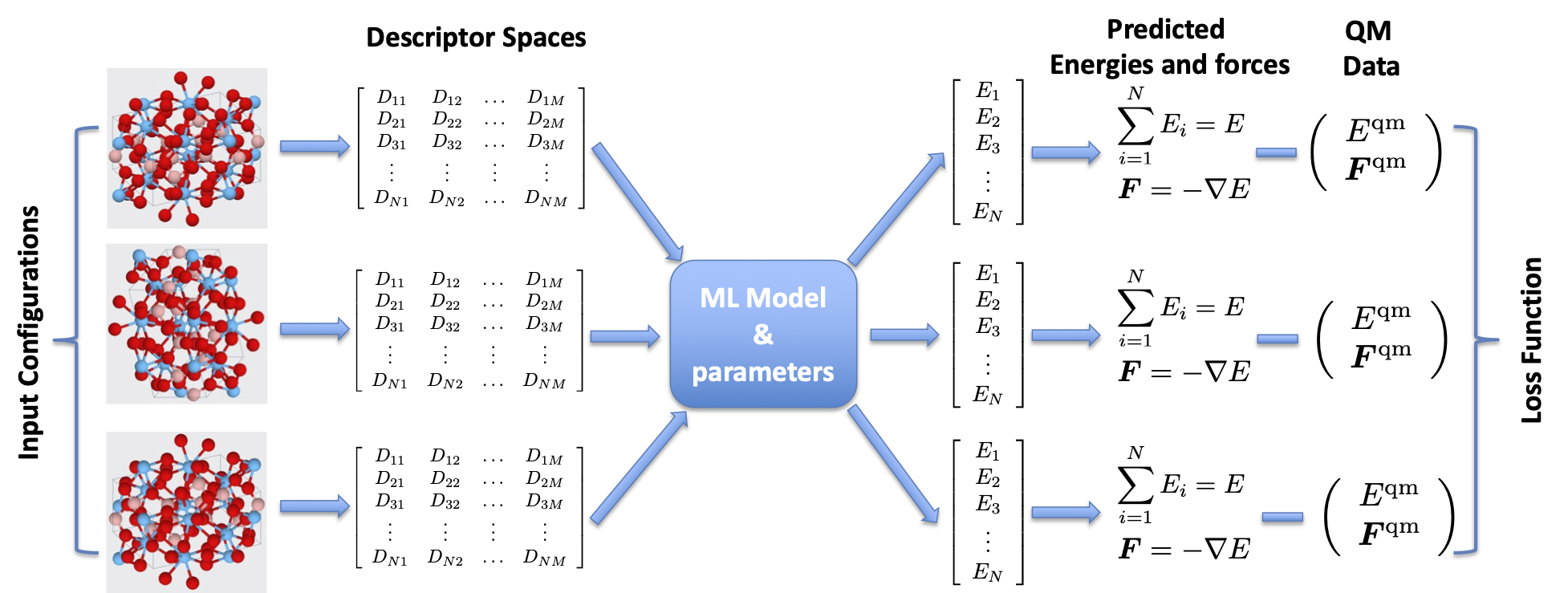} 
\caption{\label{fig2b} The composition of invariant descriptors, regression method, and QM database for building a ML model to represent the PES of a particular system. The ML model and its parameters can be defined by one of regression methods discussed in Section \ref{regressionmethods}.}
\end{figure*}

A machine learning potential is a ML model in which the inputs are atom configurations and the outputs are quantities of interest such as energies, forces, and stresses. To build ML models, there are many different methods such as linear regression, kernel regression, nonlinear regression, neural networks, and graph neural networks, which will be discussed in Section \ref{regressionmethods}. Training ML models require sufficiently large and diverse QM data sets because they often have a large number of parameters. Figure \ref{fig2b} illustrates the process of training a ML model. A crucial advantage of MLIPs is that the number of model parameters can be varied to produce several ML models with different levels of accuracy and complexity. This feature enables MLIPs to target applications requiring highly accurate prediction that can not be achieved with EIPs. ML models do not directly use Cartesian coordinates of atoms as their inputs. Instead, they require input features known as invariant descriptors which play a central role in yielding robust, accurate, transferable MLIPs. We survey a number of methods for constructing invariant descriptors and discuss their connection to the POD method. 

\subsubsection{Atom-centered symmetry functions}


The first  attempt of bringing machine learning methods to the construction of interatomic potentials  can be attributed to Behler and Parrinello who introduced the atom-centered symmetry functions (ACSFs) for constructing neural network potentials \cite{Behler2007,Behler2011,Behler2014}. The ACSF descriptors of  are derived from the radial function
\begin{equation}
\label{eq21c}
G_i^{(2)} = \sum_j e^{-\eta (r_{ij}-r_s)^2} f_{\rm c}(r_{ij})   
\end{equation}
and the angular function
\begin{equation}
\label{eq22c}
\begin{split}
G_i^{(3)} = & 2^{1 - \zeta} \sum_{j,k} (1 + \lambda \cos \theta_{ijk})^\zeta  \cdot  e^{-\eta (r_{ij}^2 + r_{ik}^2)} \\
& \times f_{\rm c}(r_{ij}) f_{\rm c}(r_{ik}) 
\end{split} 
\end{equation}
where $f_{\rm c}$ is the cut-off function. Different values of the parameters $\eta, r_s, \zeta, \lambda$ can be used to generate the ACSF descriptors. However, the ACSF descriptors are not orthogonal because they are just snapshots of the symmetry functions. 

The ACSF descriptors are invariant because they are functions of angles and distances and are summed over all possible atomic pairs and triplets within local atomic environments. Hence, they can be used as input features of neural networks to construct a ML model. Each of the three-body ACSF descriptors costs $O(N_i^2)$ per atom to evaluate using traditional implementation. Since the ACSF descriptors are similar to the three-body POD descriptors, they can be computed more efficiently using the atom density representation  presented in Section \ref{implementation}. An interesting idea is to extend the ACSF descriptors by considering the following symmetry function
\begin{equation}
\label{eq22d}
\begin{split}
G_i^{(4)} = &  \sum_{j,k,l} (1 + \alpha w_{ijk})^a (1 + \beta w_{ijl})^b (1 + \gamma w_{ikl})^c \\
& \times   e^{-\eta (r_{ij}^2 + r_{ik}^2 + r^2_{il})} f_{\rm c}(r_{ij}) f_{\rm c}(r_{ik}) f_{\rm c}(r_{il}) . 
\end{split} 
\end{equation}
This symmetry function yields a set of four-body ACSF descriptors which can significantly improve the accuracy of ACSF neural network potentials, while maintaining the linear scaling of the computational cost in terms of the number of neighbors.




\subsubsection{Permutationally  invariant polynomials}

The theory of permutationally invariant
polynomials (PIP) is based on classical invariant theory and was first introduced to fit the potential energy surfaces of small molecules \cite{Braams2009,Nguyen2018a,VanDerOord2020}. In the original implementation, each molecule had its potential energy surface defined and fit with the appropriate set of PIPs and each new molecule or molecular cluster required a completely new fit. The exponentially increasing cost of evaluating PIPs limited their applicability to molecules with a small number of degrees of freedom. The recently proposed atomic PIPs use the same polynomial basis as global PIPs but avoid the unfavorable scaling with increasing molecule size by using a cut-off distance and a truncation of the order of the expansion  \cite{VanDerOord2020}. In particular, atomic PIPs are defined as appropriate sums of products of distance and angle coordinates to make themselves  invariant with respect to permutation. It was demonstrated in \cite{VanDerOord2020} that low-dimensional PIPs can reach the same high accuracy and provide better transferability than high dimensional GAPs for a number of different materials.

Atomic PIPs and PODs are related to each other, as both use internal coordinates to construct their invariant descriptors. In particular, both use polynomials of angle coordinates. However, the radial basis functions are different for PODs and atomic PIPs. Indeed, the radial basis functions of atomic PIPs are polynomials of distance coordinates, while those of PODs are constructed from the orthogonal proper decomposition of parametrized potentials. Another difference lies in their computational complexity. Each of $q$-body PODs can be computed with $O(N_i)$, whereas each of $q$-body atomic PIPs costs $O(N_i^{q-1})$. To keep the computational cost low, only a small number of atomic PIPs is computed for each body order.  

\subsubsection{Power spectrum and bispectrum descriptors}

The starting point for constructing the power spectrum and bispectrum descriptors is to consider the neighborhood density at location $\bm r$ around a central atom $i$: 
\begin{equation}
\label{s03d}
\rho_i(\bm r) =  \sum_{j}  \delta(\bm r - \bm r_{ij}),   
\end{equation}
where $\delta$ is the Dirac-delta function. Expanding the neighbor density  in terms of the spherical harmonics $Y_{lm}(\hat{\bm{r}})$ and radial basis functions $g_k(r)$ yields
\begin{equation}
\rho_i({\bm r}) = \sum_{k=0} \sum_{l=0} \sum_{m=-l}^l c_{iklm} g_k(r) Y_{lm}(\hat{\bm{r}}),
\end{equation}
where the expansion coefficients are the inner products of the neighborhood density with $g_k(r) Y_{lm}(\hat{\bm{r}})$ as follows
\begin{equation}
\label{eq25}
 c_{iklm} = \sum_{j} g_k(r_{ij}) Y_{lm}(\hat{\bm{r}}_{ij}).
\end{equation}
The power spectrum and bispectrum descriptors are obtained as follows
\begin{equation}
\begin{split}
p_{i k_1k_2l} & = \sum_{m=-l}^l ({c}_{ik_1lm})^*  c_{ik_2lm} ,     \\
b_{i k_1 k_2 l_1l_2l_3} & = \sum_{m_1=-l_1}^{l_1} \sum_{m_2=-l_2}^{l_2} \sum_{m_3=-l_3}^{l_3}  ({c}_{ik_1l_1m_1})^* \\
& \qquad \qquad C^{l_1l_2l_3}_{m_1m_2m_3} c_{i k_2 l_2 m_2} c_{i k_2 l_3 m_3} ,
\end{split}
\end{equation}
where $C^{l_1l_2l_3}_{m_1m_2m_3}$ are the Clebsch-Gordan coefficients and the asterisk denotes the complex conjugation \cite{Bartok2013}. 

The power spectrum descriptors are three-body, while the bispectrum descriptors are four-body. An advantage of the power spectrum and bispectrum descriptors is that their evaluation scales linearly with the number of neighbors. Since the power spectrum and bispectrum descriptors can be expressed as sums of products of polynomials of $w_{ijk}$ and radial basis functions \cite{Bartok2013}, they are related to the fast POD descriptors.

\subsubsection{Smooth overlap of atomic positions}

Instead of the Dirac-delta functions, the SOAP method \cite{Bartok2013} constructs the neighborhood density using Gaussians expanded in terms of spherical harmonic
functions as follows
\begin{equation}
\label{s03e}
\rho_i(\bm r) =  \sum_{j} \exp(- \alpha |\bm r - \bm r_{ij}|^2) .
\end{equation}
The SOAP descriptors are equivalent to using the  power or bispectrum descriptors together with Gaussian atomic neighbor density contributions and a dot product covariance kernel. The Gaussian width controls  the smoothness of the similarity measure among atomic neighbor environments. The ability to controlling the smoothness of the similarity measure is an advantage of SOAP over the  power or bispectrum descriptors \cite{Bartok2017,Bartok2018,Deringer2017,Dragoni2018}.

\subsubsection{Spectral neighbor analysis potential}

The SNAP descriptors \cite{Thompson2015} are constructed from hyperspherical harmonics on the unit 3-sphere
\begin{equation}
\label{4dlap1}
U_{jmm'}(\omega, \phi, \theta) = (\sin \omega)^m G_{j-m}^{m+1}(\cos \omega) Y_{mm'}(\theta,\phi)
\end{equation}
with
\begin{equation*}
\phi = \arctan (y/x), \quad \theta = \arccos(z/|\bm r|), \quad  \omega = \pi |\bm r|/ r_0 .
\end{equation*}
Here $G_{j-l}^{l+1}$ are the Gegenbauer polynomials, and $Y_{mm'}$ are the usual spherical harmonics. These hyperspherical harmonics are also known as the Wigner matrices \cite{Thompson2015}. 
To make the contribution from atoms at $r = r_{\rm cut}$
vanish smoothly to zero, it is necessary to augment the atomic
neighbor density function (\ref{s03d}) with a cutoff function
\begin{equation}
\label{4dd}
\rho_i(\bm r) = \sum_{j}  f_{\rm c}(r _{i})  \delta (\bm r - \bm r_{ij}) .
\end{equation}
This atomic neighborhood density is expanded in terms of hyperspherical harmonics on the unit 3-sphere as
\begin{equation}
\rho_i(\bm r) = \sum_{j=0, 1/2,\ldots}^\infty \sum_{m=-j}^j \sum_{m'=-j}^j c_{ijmm'}  U_{jmm'}(\omega, \phi, \theta)
\end{equation}
where
\begin{equation}
 c_{ijmm'}  =  \sum_{j} f_{\rm c}(r _{ij})  U_{jmm'}(\omega_{ij}, \phi_{ij}, \theta_{ij}) .
\end{equation}
The coefficients  $c_{ijmm'}$ are used to define the following bispectrum components
\begin{equation}
\label{snapbi}
\begin{split}
b_{ij_1j_2j} & =  \sum_{m_1,m_1'=-j_1}^{j_1} \sum_{m_2,m_2'=-j_2}^{j_2} \sum_{m,m'=-j}^{j}  \left({c}_{ijmm'} \right)^* \\
& \qquad C^{j j_1j_2}_{m m_1 m_2} C^{j j_1 j_2}_{m' m'_1 m'_2} \ c_{i j_1m_1m'_1} c_{i j_2 m_2m_2'} ,    
\end{split}
\end{equation}
where $C^{j j_1j_2}_{m m_1 m_2}$ are the Clebsch–Gordan coefficients. We see from (\ref{4dlap1}) that the hyperspherical harmonics can be expressed as {the product of a radial function and a spherical harmonic}  $U_{jmm'}(\omega, \phi, \theta) = g_{k}(r) Y_{mm'}(\theta, \phi) $ with $g_{k}(r) = (\sin \omega)^m G_{j-m}^{m+1}(\cos \omega)$, where the index $k$ has a one-to-one correspondence to $(j,m)$. {Connection between the SNAP descriptors and the spherical harmonics basis combined with a radial basis of Gegenbauer polynomials has been noted in \cite{Drautz2020,Lysogorskiy2021}.}  SNAP can be viewed as a variant of the bispectrum descriptors and thus related to the four-body POD descriptors.


\subsubsection{Atomic cluster expansion}


The atomic cluster expansion (ACE) is recently developed in \cite{Drautz2019,Drautz2020} as a hierarchical and complete descriptor of the local atomic environment due to its arbitrary number of body orders. This method extends the power and bispectrum descriptors to obtain a complete set of invariant descriptors. The ACE method encompasses several well-known descriptors since ACSF, SOAP, power spectrum, bispectrum, and SNAP can be expanded using the ACE descriptors. In the ACE method, the atomic basis functions are defined as follows
\begin{equation}
B_{inlm} = \sum_{j=1}^{N_i} R_{nl}(r_{ij}) Y_{lm} (\hat{\bm r}_{ij})
\end{equation}
where the radial basis functions $R_{nl}(r)$ are constructed from the Chebyshev polynomials of the first
kind and the cosine cut-off function, and $Y_{lm}$ are complex spherical harmonics. The ACE descriptors are then computed as
\begin{equation}
\mathcal{D}^{(2)}_{in} = B_{in00}
\end{equation}
\begin{equation}
\mathcal{D}^{(3)}_{in_1n_2l} = \sum_{m=-l}^l (-1)^m B_{i n_1l m} B_{i n_2l -m}
\end{equation}
\begin{equation}
\mathcal{D}^{(4)}_{\substack{in_1n_2n_3\\ l_1 l_2 l_3}} = \sum_{m_1m_2m_3}  W_{m_1m_2m_3}^{l_1l_2l_3} B_{i n_1l_1 m_1} B_{i n_2l_2 m_2} B_{i n_3l_3 m_3}
\end{equation}
which are in that order the two-body, three-body, and four-body, respectively. The coupling constants $W_{m_1m_2m_3}^{l_1l_2l_3}$ are calculated from the Clebsch–Gordan coefficients. The paper \cite{Drautz2019} describes a systematic approach to defining many-body ACE descriptors. The ACE descriptors are complete in the sense that they can converge to cluster functional to arbitrary accuracy. They scale linearly with the number of neighbors irrespective of the order of the expansion. 

The POD method bears similarities with the ACE method. We note that the formulae of the atomic basis functions for the ACE method and our method are similar. However, the two methods employ different radial and angular basis functions to define the atomic basis functions. While the ACE method uses complex spherical harmonics $Y_{lm} (\hat{\bm r}_{ij})$, our method uses angular monomials $A_{lm} (\hat{\bm r}_{ij})$. An advantage of our method is that the angular monomials are less expensive to compute than the spherical harmonics. While the ACE method uses the Chebyshev polynomials of the first kind, our method uses the proper orthogonal decomposition to construct the orthogonal radial basis functions from parametrized potentials. The number of ACE descriptors and thus the computational cost grow exponentially with the body orders in terms the number of radial basis functions, whereas those of our method grow linearly with the body orders in terms of the number of radial basis functions.


\subsubsection{Moment tensor potentials}

The moment tensor potentials (MTP) first proposed in \cite{Shapeev2016} are a polynomial expansion of both the radial and angular components. The basis set of descriptors is very similar to the one of atomic cluster expansion (ACE) \cite{Drautz2019,Drautz2020} and related to the permutation-invariant polynomial (PIP) basis \cite{Braams2009,Nguyen2018a,VanDerOord2020}.  The descriptors of the moment tensor potentials are obtained by contraction of so-called moment tensors, which are defined as
\begin{equation}
M_{n,q} = \sum_{j=1}^{N_i} R_{n}(r_{ij}) \underbrace{\bm r_{ij} \otimes  \bm r_{ij} \otimes  \ldots \otimes  \bm r_{ij}}_{q \ \mbox{times}} 
\end{equation}
where the radial part $R_{n}(r_{ij})$ consists of Chebyshev polynomials and the angular part $\bm r_{ij} \otimes  \bm r_{ij} \otimes  \ldots \otimes  \bm r_{ij}$ contains angular information about the neighborhood. If $q = 2$ the angular part has the form of the matrix
\begin{equation}
\bm r_{ij} \otimes  \bm r_{ij} = \left(
\begin{array}{ccc}
 x_{ij}^2  & x_{ij} y_{ij}  & x_{ij} z_{ij} \\
x_{ij} y_{ij}  &  y_{ij}^2 & y_{ij} z_{ij} \\
x_{ij} z_{ij}  & y_{ij} z_{ij}  & z_{ij}^2 
\end{array}
\right) .
\end{equation}
To achieve rotational invariance the contraction of the moment tensors must produce reducible representations of the identity representation of the rotation group. Such contraction of the moment tensors is described in detail in \cite{Shapeev2016}.

We see that moment tensors are formed from the products of the radial basis functions and the angular monomials. In other words, the moment tensors and the atomic basis functions of our method span the same approximation space if the same radial basis functions are used.  Furthermore, the tensor contraction of the moment tensor potentials can be expressed as the sum of the products of the atomic basis functions. An advantage of PODs is that the angular basis functions require less computational cost and memory storage than the moment tensors.  Another difference between PODs and the moment tensor potentials lies in the choice of radial basis functions.

\subsubsection{Atom density fields versus internal coordinates}

As articulated in \cite{Musil2021}, all existing descriptors are constructed from Cartesian coordinates by using either atom density fields or internal coordinates. A crucial advantage of the atom density approach is that with an appropriate discretization of the angular basis one can evaluate atom density descriptors as a sum of products of the density coefficients. This ensures that the cost of computing all descriptors of a given body order scales only linearly with the number of neighbors, even though it scales exponentially with the body order in terms of the number of basis functions. In contrast, the cost of ACSFs and PIPs depends linearly on the number of basis functions but exponentially with the body order in terms of the number of neighbors because their descriptors are evaluated over all possible tuples composed of neighbors of the central atom. This crucial difference makes atom density methods more efficient when there are many neighbors and the body orders are high. Despite this rather fundamental difference in philosophy and computational cost, the two families of representations compute entities that are essentially equivalent.

In this paper, we show that it is possible to compute certain classes of internal coordinate descriptors with linear scaling cost if they can be recast into atom density fields. We show how to achieve that for fast PODs in Section \ref{implementation}, but the same trick can be used to bring the cost of both ACSFs and atomic PIPs to scale linearly with the number of neighbors. More generally, internal coordinate descriptors of the form
\begin{equation}
\mathcal{D}^{(q+1)}_{im} = \sum_{j_1, \ldots, j_q} g_m(\bm r_{ij_1}, \ldots, \bm r_{ij_q}) 
\end{equation}
can be reformulated into atom density descriptors. However, we emphasize that there are classes of internal coordinate descriptors that can not be recast into atom density fields. For instance, the following bond-order descriptors can not be reformulated as sums of the products of atomic basis functions 
\begin{equation}
\label{eq58b}
\mathcal{D}_{inm}^{(3)} = \sum_{j=1}^{N_i} h_n(r_{ij}) f \left(\sum_{k=1}^{N_i} g_m(r_{ij}, r_{ik}, w_{ijk}) \right)  
\end{equation}
if $f$ is a nonlinear function. As a result, these descriptors have to be computed with $O(N_i^2)$. The descriptors (\ref{eq58b}) can be instrumental to constructing bond order potentials discussed in Section \ref{Empirical}. 

The space of internal coordinate descriptors encompasses the space of atom density descriptors. This is because all classes of atom density descriptors can be expressed as functions of internal coordinates, while only certain classes of internal coordinate descriptors can be recast into atom density descriptors. Internal coordinate descriptors are more intuitive than atom density descriptors because, like empirical potentials, they are built directly from distances and angles. Indeed, internal coordinate descriptors are nothing but a hierarchical collection of empirical potentials which are constructed by using systematic frameworks. Following ACSF's framework, one can generate the SW descriptors or the Tersoff  descriptors  by evaluating the SW potential or the Tersoff potential at different values of their parameters. One can also use PIP or POD frameworks to generate invariant descriptors as functions of distances and angles, which are symmetrized to make themselves invariant with respect to permutation. In the next section, we discuss several methods that compose any set of invariant descriptors to construct PESs.

\section{Regression methods}
\label{regressionmethods}

Regression method is another crucial ingredient in the making of interatomic potentials. Regression method combines descriptors or functions of descriptors in a mathematical model to represent the PES of a particular system as illustrated in Figure \ref{fig2b}. Once the model is fitted against QM database, it can be used to predict physical properties and perform MD simulations of the system. In this section, we discuss commonly used regression methods for building interatomic potentials.

\subsection{Linear regression}

Linear regression is the most simple and efficient method. The atomic energy at an atom $i$ is expressed as a linear combination of the local descriptors 
\begin{equation}
\label{eq57a}
E_i = \sum_{m=1}^M c_m \mathcal{D}_{im}   
\end{equation}
where ${\mathcal{D}}_{im}$ are the local descriptors at atom $i$ and $c_m$ are the coefficients to be determined by fitting against QM database. Here $M$ is the number of descriptors per atom. Thus, the PES is given by
\begin{equation}
\label{eq58a}
E = \sum_{i=1}^N E_i = \sum_{m=1}^M c_m G_{m}     
\end{equation}
where $G_m = \sum_{i=1}^N \mathcal{D}_{im}$ are the global descriptors. The coefficients $c_m$ are sought as solution of a least squares problem 
\begin{equation}
\label{eq63a}
\min \|\bm A \bm c - \bm b \|^2 + \gamma \|\bm c\|^2 
\end{equation}
where the matrix $\bm A$ is formed from the global descriptors and their derivatives for many atom configurations in the training database, the vector $b$ is comprised of QM data such as DFT energies, forces, and stresses. Note that $\gamma$ is a regularization parameter that helps stabilize the linear regression especially if the matrix $\bm A$ is not well-conditioned. In \cite{Wood2018}, the energy expansion (\ref{eq57a}) is extended to
\begin{equation}
E_i = \sum_{m=1}^M c_m \mathcal{D}_{im} +  \sum_{m=1}^M \sum_{n=m}^M \beta_{mn} \mathcal{D}_{im}  \mathcal{D}_{in} .  
\end{equation}
The resulting model is linear in the coefficients and quadratic in the descriptors. The quadratic terms $\mathcal{D}_{im}  \mathcal{D}_{in}$ increase both the size and the body order of the resulting model relative to the original model (\ref{eq57a}). It is shown in \cite{Wood2018} that the quadratic terms increase the accuracy of the model significantly.


Linear regression has been widely used to construct interatomic potentials \cite{Thompson2015,Wood2018,Novoselov2019,Shapeev2016}. Beside the beauty of simplicity  and efficiency, linear regression has an advantage for uncertainty quantification. Owing to the linearity of the model (\ref{eq58a}), both the mean and covariance of the PES can be computed efficiently when the coefficients $c_m$ are treated as random variables with known distributions. This trait of linear regression has been exploited to quantify prediction errors and develop active learning algorithms \cite{Musil2021,Novikov2021,Podryabinkin2017}. 

\subsection{Kernel regression}

Gaussian approximation potentials (GAPs) \cite{Bartok2010,Fujikake2018,Szlachta2014,Deringer2021} are an important example for kernel regression, which are obtained by combining a suitable structural descriptor and a kernel establishing the connection between structure and energy.
\begin{equation}
\label{eq61a}
E_i = \sum_n^{N_{\rm train}} \alpha_n G(\bm{\mathcal{D}}'_{n}, \bm{\mathcal{D}}_i)
\end{equation}
where $N_{\rm train}$ is the total number of atoms in the training set and $\bm{\mathcal{D}}'_{n}$ is the vector of descriptors for atom $n$ in the training set. GAPs use the squared exponential kernel
\begin{equation}
\label{eq62a}
G(\bm{d}', \bm{d}) = \exp \left( - \sum_{m=1}^M \frac{(d'_m - d_m)^2}{\theta_m} \right),
\end{equation}
 where $\theta_m$ are hyper parameters. The determination
of the parameters $\alpha_n$ requires an inversion of the
covariance matrix of size $N_{\rm train} \times N_{\rm train}$. For many applications, $N_{\rm train}$ can exceed $10^5$, and therefore GAPs can be computationally expensive. To reduce the computational cost, one can choose representative atoms from the full set of training atoms to construct GAPs \cite{Deringer2021}.

As a statistical machine learning technique, GAPs can compute not only the mean of the prediction, but also the variance which provides an error estimate for the prediction. Active learning for GAPs was demonstrated for liquid and amorphous
phases of hafnium dioxide \cite{Sivaraman2020}, and melting and point defect diffusion in aluminum \cite{Jinnouchi2019}.


\subsection{Nonlinear regression}


Nonlinear regression is prominently used to fit empirical potentials \cite{Daw1984,Baskes1992,Stillinger1985,Tersoff1988,Bazant1997,Brenner2002,VanDuin2001}. Typically, it involves minimizing a loss function which is the sum of the errors between the model and QM calculation as follows 
\begin{equation}
\label{eq63b}
\begin{split}
\min_{\bm \beta} \sum_{k=1}^{N_{\rm config}} \Big[ w^e_k \left(E(\bm R_k; \bm \beta) - E^{\rm qm}(\bm R_k) \right)^2 \\
+ w^f_k  \left\|\bm F(\bm R_k; \bm \beta) - \bm F^{\rm qm}(\bm R_k) \right\|^2 \Big]
\end{split}
\end{equation}
where $\bm \beta$ is a vector of parameters to be optimized, $\bm R_k$ is a $k$th configuration in the training set, $E(\bm R_k; \bm \beta)$ and $\bm F(\bm R_k; \bm \beta)$ are the model energie and forces, $E^{\rm qm}(\bm R_k)$ and $\bm F^{\rm qm}(\bm R_k)$ are QM energy and forces. The weights $w_k^e$ and $w_k^f$ are chosen to place the relative importance of energies and forces for every configurations. Recently, an efficient framework \cite{Bochkarev2022} for parametrization of ACE models employs nonlinear regression to fit a PES of the form
\begin{equation}
\label{eq571a}
E_i = F \left( \sum_{m=1}^M \alpha_{m} \mathcal{D}_{im} \right) + \sum_{m=1}^M \beta_{m} \mathcal{D}_{im} .  
\end{equation}
with a general nonlinear embedding function $F$. The nonlinear ACE models are optimized by using Tensor Flow to solve the problem (\ref{eq63b}). It is shown in \cite{Bochkarev2022} that nonlinear ACE models provide more accurate predictions than linear ACE models for the same computational cost.

\subsection{Neural network regression}

Neural network regression is a generalization of nonlinear regression. The descriptor vectors $\bm{\mathcal{D}}_i$ are input features for a neural net. For simplicity of exposition we assume that the neural net has $L$ hidden layers and the number of neurons for each hiden layers is equal to the number of descriptors per atom. Starting with $\bm{\mathcal{D}}^0_i = \bm{\mathcal{D}}_i$, the descriptors at the hidden layer $l$ are given by 
\begin{equation}
\label{eq671a}
\bm{\mathcal{D}}^{l}_i = \bm \sigma( \bm W^{l} \bm{\mathcal{D}}_{i}^{l-1} + \bm b^l),  \quad l = 1, \ldots, L,
\end{equation}
where $\bm \sigma$ are non-linear activation functions such as the hyperbolic tangent or Gaussian function, $\bm W^l$ and $\bm b^l$ are weight matrices and bias vectors.
The atomic energy at an atom $i$ is expressed as a linear combination of the descriptors at the last hidden layer as
\begin{equation}
\label{eq672a}
E_i = \sum_{m=1}^M c_m \mathcal{D}_{im}^L .  
\end{equation}
Forces are calculated by using reverse differentiation as
\begin{equation}
\label{eq673a}
\frac{d E_i}{d \bm R_i} = \frac{\partial E_i}{\partial \bm{\mathcal{D}}^{L}_i} \frac{\partial \bm{\mathcal{D}}^{L}_i}{\partial \bm{\mathcal{D}}^{L-1}_i} \ldots \frac{\partial \bm{\mathcal{D}}^{1}_i}{\partial \bm{\mathcal{D}}^{0}_i} \frac{\partial \bm{\mathcal{D}}^{0}_i}{\partial \bm R_i}
\end{equation}
where $\frac{\partial E_i}{\partial \bm{\mathcal{D}}^{L}_i}  = \bm c$, $\frac{\partial \bm{\mathcal{D}}^{l}_i}{\partial \bm{\mathcal{D}}^{l-1}_i} =  \bm \sigma'( \bm W^{l} \bm{\mathcal{D}}_{i}^{l-1} + \bm b^l) \bm W^{l}$, and $\bm \sigma'$ are first derivatives of $\bm \sigma$. Note that $\frac{\partial \bm{\mathcal{D}}^{0}_i}{\partial \bm R_i}$ are partial derivatives of the input descriptors with respect to  positions of neighbors of atom $i$. All the parameters of the neural net are stacked into a vector $\bm \beta$ which is found by solving the minimization problem (\ref{eq63b}). 

In \cite{Blank1995}, neural networks have first been used to fit PESs from at a limited number of configurations. Subsequently, a large number of NN potentials have been developed for many atomic and molecular systems \cite{Artrith2016,Behler2007,Behler2011,Behler2014,Behler2021,Wang2018}.


\subsection{Graph neural network regression}


Graph neural network is a generalization of neural net. We  describe graph convolution layers \cite{Schutt2018,Schutt2020} which are one of the simplest graph neural networks.  Let $\bm A \in \mathbb{R}^{N \times N}$ be the adjacency matrix for a particular configuration $\bm R$, that is, $A_{ij} = 1$ if $|\bm r_{i} - \bm r_{j}| \le r_{\rm cut}$ and $i \neq j$, otherwise $A_{ij} = 0$ .  Let $\bm{\mathcal{D}} \in \mathbb{R}^{N \times M}$ be the  descriptor matrix whose row $i$ consists of the $M$ local descriptors for atom $i$. Starting from $\bm{\mathcal{D}}^0 = \bm{\mathcal{D}}$, the graph convolution layers consist of the following propagation  
\begin{equation}
\label{eq68a}
\bm{\mathcal{D}}^{l} =  \bm  \sigma \left( \bm A \bm{\mathcal{D}}^{l-1}  \bm W^l \right),    \quad l = 1, \ldots, L .
\end{equation}
The PES of the graph convolution layers is expressed as
\begin{equation}
E = \bm e^T \bm{\mathcal{D}}^{L} \bm c
\end{equation}
where $\bm e^T = (1, 1, \ldots, 1)$ and $\bm c$ is a vector coefficients at the output layer. Differentiating (\ref{eq68a}) with respect to the atom positions yields 
\begin{equation}
\frac{\partial \bm{\mathcal{D}}^{l}}{\partial \bm R} =   \bm \sigma' (\bm A \bm{\mathcal{D}}^{l-1}  \bm W^l)   \bm A  \frac{\partial \bm{\mathcal{D}}^{l-1}}{\partial \bm R}  \bm W^l, 
\end{equation}
where $\bm \sigma'$ denotes the derivative of the activation function. Hence, the forces are computed as
\begin{equation}
\bm F = - \bm e^T \frac{\partial \bm{\mathcal{D}}^{L}}{\partial \bm R} \bm c .
\end{equation}
 In practice, since the multiplication with $\bm A$ completely changes the magnitude of the input features at the next layer, $\bm A$ is replaced with $\hat{\bm D}^{-1/2} \hat{\bm A} \hat{\bm D}^{-1/2}$, where $\hat{\bm A} = \bm A + \bm I$ and $\hat{\bm D}$ is the degree matrix of $\hat{\bm A}$, to maintain the magnitude of the input features across all layers. The training of GNN models is similar to that of NN models.

The development of GNN potentials is an emerging trend in machine learning interatomic potentials. A  number of GNN potentials have been recently developed, including SchNet \cite{Schutt2018,Schutt2020}, DimeNet, \cite{NEURIPS2021_82489c97}, GemNet \cite{NEURIPS2021_35cf8659}, and NequIP \cite{Batzner2022}. Although these GNN potentials have rather complex architectures, many of them employ the graph convolution layers as their building blocks. Thus far, little research has been conducted to show the application of GNN potentials in real atomistic simulations. In the recent paper \cite{Batzner2022}, NequIP was demonstrated to produce stable and accurate MD simulations. More recently, the paper \cite{Stocker2022} provides an in-depth exploration of the robustness of GemNet in MD simulations. 

\section{Results and Discussions}
\label{results}

The fast POD method will be demonstrated and compared with SNAP method \cite{Thompson2015} and the original POD method \cite{Nguyen2022} for Tantalum element. Linear regression (\ref{eq63a}) is used with $\gamma = 10^{-12}$ to construct interatomic potentials using SNAP,  original POD, and fast POD descriptors. The models are fitted against DFT energies and forces from a QM data set available in the FitSNAP package. Here the fitting weights for energies and forces are set to 100 and 1, respectively. The inner and outer cut-off distances are set to $r_{\rm in} = 1$\AA \ and $r_{\rm cut} = 5$\AA, resepectively. We fit the SNAP models with the FitSNAP package, and the original POD models with ML-POD package  available in LAMMPS \cite{Thompson2022}. Both the source code and the fast POD models are available upon request to facilitate the reproduction of our work. 


\subsection{Results for Tantalum}

The Ta data set contains a wide range of configurations to adequately sample the important regions of the potential energy surface \cite{Thompson2015}. The data set includes 12 different groups such as surfaces, bulk structures, defects, elastics for BCC, FCC, and A15 crystal structures, and high temperature liquid. The database was used to create a SNAP potential \cite{Thompson2015} which successfully describes a wide range of properties such as energetic properties of solid tantalum phases, the size and shape of the Peierls barrier for screw dislocation motion in BCC tantalum, as well as both the structure of molten tantalum and its melting point. Herein  the database is used to fit SNAP, POD, and fast POD potentials. We consider two different fast POD potentials: the first fast POD potential (FPOD3) is a three-body potential that has the same number of descriptors as the original POD potential (OPOD); the second fast POD potential (FPOD4) is based on the first fast POD potential and added with four-body descriptors. The purpose of FPOD3 is  to demonstrate the performance of the fast POD method relative to the original POD method, while the porpose of FPOD4 is to demonstrate the accuracy gain obtained by adding the four-body POD descriptors to FPOD3. 

Table \ref{table6} displays the number of descriptors for all potentials. Note that all potentials have a one-body descriptor to account for isolated energies. We purposely choose the numbers of descriptors for OPOD and FPOD3 models to match those for SNAP models. FPOD4 models have more descriptors than the other models because they are augmented with additional four-body POD descriptors. We evaluate these models on the basis of mean absolute errors (MAEs) of the predicted energies and forces, which are defined as follows
\begin{equation}
\begin{split}
\varepsilon_E & = \frac{1}{N_{\rm config}} \sum_{n = 1}^{N_{\rm config}} |E_n - E_n^{\rm qm}| \\
\varepsilon_F & = \frac{1}{N_{\rm force}} \sum_{n = 1}^{N_{\rm force}} |F_{n} - F_{n}^{\rm qm}| \\
\end{split}
\end{equation}
where $N_{\rm config}$ is the number of configurations in a data set, and $N_{\rm force}$ is the total number of force components for all the configurations in the same data set. 

\begin{table}[htbp]
\caption{\label{table6}%
Number of descriptors $M$ for SNAP, OPOD, FPOD3, and FPOD4. For SNAP, $J$ is the upper limit for the index $j$ of $U_{jmm'}(\omega, \phi, \theta)$ in (\ref{4dlap1}) and $M = 1 + (J+1)(J+2)(J+3/2)/3$. For OPOD and FPOD3, $M = 1 + K_r^{(2)} + K_r^{(3)}(1+ P_a^{(3)})$. For FPOD4, $M = 1 + K_r^{(2)} + K_r^{(3)}(1+ P_a^{(3)}) + K_r^{(4)}K_a^{(4)}$.
}
\begin{ruledtabular}
\begin{tabular}{cc|cccc|cccc}
\multicolumn{2}{c|}{\textbf{SNAP}} &
\multicolumn{4}{c|}{\textbf{OPOD/FPOD3}} & 
\multicolumn{4}{c}{\textbf{FPOD4}} \\   
\hline
$J$ & $M$ & $K_r^{(2)}$ & $K_r^{(3)}$ & $P_a^{(3)}$  & $M$ & $K_r^{(4)}$ & $P_a^{(4)}$ & $K_a^{(4)}$ & $M$  
 \\
\hline
1 &  6 & 2 & 1 & 2 & 6 & 1 & 1 & 1 & 8   \\
2 & 15 & 5 & 3 & 2 & 15 & 3 & 2 & 2&  27   \\
3 & 31 & 5 & 5 & 4 & 31 & 5 & 3 & 4 & 66  \\
4 & 56 & 6 & 7 & 6 & 56 & 7 & 4 & 7 & 133  \\
5 & 92 & 10 & 9 & 8 & 92 & 9 & 5 & 16 & 236  \\
6 & 141 & 8 & 12 & 10 & 141 & 12 & 6 & 23 & 417\\
\end{tabular}
\end{ruledtabular}
\end{table}

\begin{table}[htbp]
\caption{\label{table7}%
Training errors in energies and forces for SNAP, OPOD, FPOD3, FPOD4 potentials listed in Table \ref{table6}. The units of the mean absolute
errors of the energy and force are meV/atom and meV/\AA, respectively.
}
\begin{ruledtabular}
\begin{tabular}{cc|cc|cc|cc}
\multicolumn{2}{c|}{\textbf{SNAP}} &
\multicolumn{2}{c|}{\textbf{OPOD}} &
\multicolumn{2}{c|}{\textbf{FPOD3}} &
\multicolumn{2}{c}{\textbf{FPOD4}} \\   
\hline
$\varepsilon_E$ & $\varepsilon_F$ & $\varepsilon_E$ & $\varepsilon_F$ & $\varepsilon_E$ & $\varepsilon_F$ & $\varepsilon_E$ & $\varepsilon_F$  \\
\hline 
494.01 &   636.88 &   137.49 &   269.39 &   107.18 &   179.38 &   106.12 &   185.01 \\
76.55 &   257.56 &    44.31 &   331.50 &    62.23 &   263.23 &    33.95 &   234.32 \\ 
14.50 &   138.16 &     8.50 &   116.44 &     8.67 &   139.73 &     3.86 &   102.54 \\ 
6.66 &    90.14 &     4.06 &    98.82 &     4.97 &   105.95 &     1.22 &    55.18 \\ 
3.26 &    72.92 &     1.77 &    67.83 &     1.79 &    68.31 &     0.77 &    42.04  \\
1.41 &    68.10 &     1.05 &    53.93 &     1.03 &    54.91 &     0.48 &    34.72 
\\
\end{tabular}
\end{ruledtabular}
\end{table}


Table \ref{table7} show training errors in energies and forces predicted for SNAP, OPOD, FPOD3, FPOD4 potentials for different values of the number of the descriptors listed in Table \ref{table6}. As expected, both the energy and force errors decrease as the number of descriptors increases. Note that the convergence of the energy errors is significantly faster than that of the force errors for all potentials. While the energy errors drop more than two orders of magnitude, the force errors drop less than one order of magnitude. As expected, both OPOD and FPOD3 have very similar energy and force errors, because the three-body descriptors of OPOD in (\ref{eq15b}) are very similar to those of FPOD3 in (\ref{eq15}), while their two-body descriptors are exactly the same. We observe that both OPOD and FPOD3 yield faster convergence of the errors than SNAP. As expected, FPOD4 yields faster convergence in energy and force errors than both OPOD and FPOD3 owing to the fact that it has the additional four-body descriptors. The MAEs in energies reach 1.41 meV/atom for SNAP, 1.05 meV/atom for OPOD, 1.03 meV/atom for FPOD3, and 0.48 meV/atom for FPOD4. These energy errors are generally below the limits of DFT errors. The MAEs in forces reach  68.10 meV/\AA \ for SNAP, 53.93 meV/\AA \ for OPOD, 54.91 meV/\AA \ for FPOD3, and 34.72 meV/\AA \ for FPOD4. These force errors are acceptable for most applications. Smaller errors can be achieved by further increasing the number of descriptors albeit at the expense of increasing the computational cost.

Table \ref{table8} provides the training errors in energy and forces for each of the 12 groups. The Surface group has the the highest mean absolute error in energy, while the Liquid group has the highest mean absolute error in forces. The liquid structures depend strongly on the repulsive interactions that occur when two atoms approach each other. Consequently, it is more difficult for potential models to predict forces of the liquid structures since  the liquid configurations are very different from those of the equilibrium solid crystals.  It is also more difficult to predict energies of surface configurations because the surfaces of BBC crystals tend to be rather open with surface atoms exhibiting rather low coordination numbers. The MAE force errors for Volume A15, Volume BCC, and Volume FCC are zero, because all of the structures in Volume A15, Volume BCC, and Volume FCC are at equilibrium states and thus have zero atomic forces. FPOD4 has smaller energy and force errors than both SNAP and OPOD across all training groups. 
 
\begin{table}[htbp]
\caption{\label{table8}%
Energy and force errors for SNAP with $M = 56$, OPOD with $M = 56$, FPOD3 with $M = 56$, FPOD4 with $M = 133$ for different training groups in the data set. The units of the mean absolute errors of the energy and force are meV/atom and meV/\AA, respectively.
}
\begin{ruledtabular}
\begin{tabular}{c|cc|cc|cc}    
    Training  & \multicolumn{2}{c|}{\textbf{SNAP}} & \multicolumn{2}{c|}{\textbf{OPOD}} & 
     \multicolumn{2}{c}{\textbf{FPOD4}}\\    
    group & $\varepsilon_E$ & $\varepsilon_F$ & $\varepsilon_E$ & $\varepsilon_F$ & $\varepsilon_E$ & $\varepsilon_F$  \\
\hline 
Displaced A15 &7.68 &   107.99 &     1.10 &   112.90 &     0.56 &    62.03 \\ 
Displaced BCC &18.54 &   165.20 &     2.38 &   195.78 &     0.78 &   109.95 \\ 
Displaced FCC & 0.53 &    63.10 &     4.33 &   115.39 &     2.26 &    48.20 \\ 
Elastic BCC & 0.29 &     0.09 &     0.67 &     0.06 &     0.27 &     0.03 \\ 
Elastic FCC & 0.88 &     0.12 &     0.82 &     0.11 &     0.33 &     0.10 \\ 
GSF 110 & 18.67 &    25.56 &     5.19 &    38.74 &     1.21 &    25.82 \\ 
GSF 112 & 7.47 &    99.37 &     6.42 &    67.78 &     2.86 &    46.85 \\ 
Liquid & 18.33 &   395.62 &     6.40 &   433.97 &     1.76 &   229.60 \\ 
Surface & 21.96 &    55.82 &    11.15 &    50.64 &     8.01 &    40.69 \\ 
Volume A15 & 15.66 &          0 &   7.46 &          0   & 2.53 &          0\\
Volume BCC & 19.03 &          0  & 12.40 &          0   & 3.15 &          0\\
Volume FCC & 13.11 &          0  & 13.43 &          0   & 1.86 &          0\\
\end{tabular}
\end{ruledtabular}
\end{table}

\begin{figure}[htbp]
\includegraphics[scale=0.425]{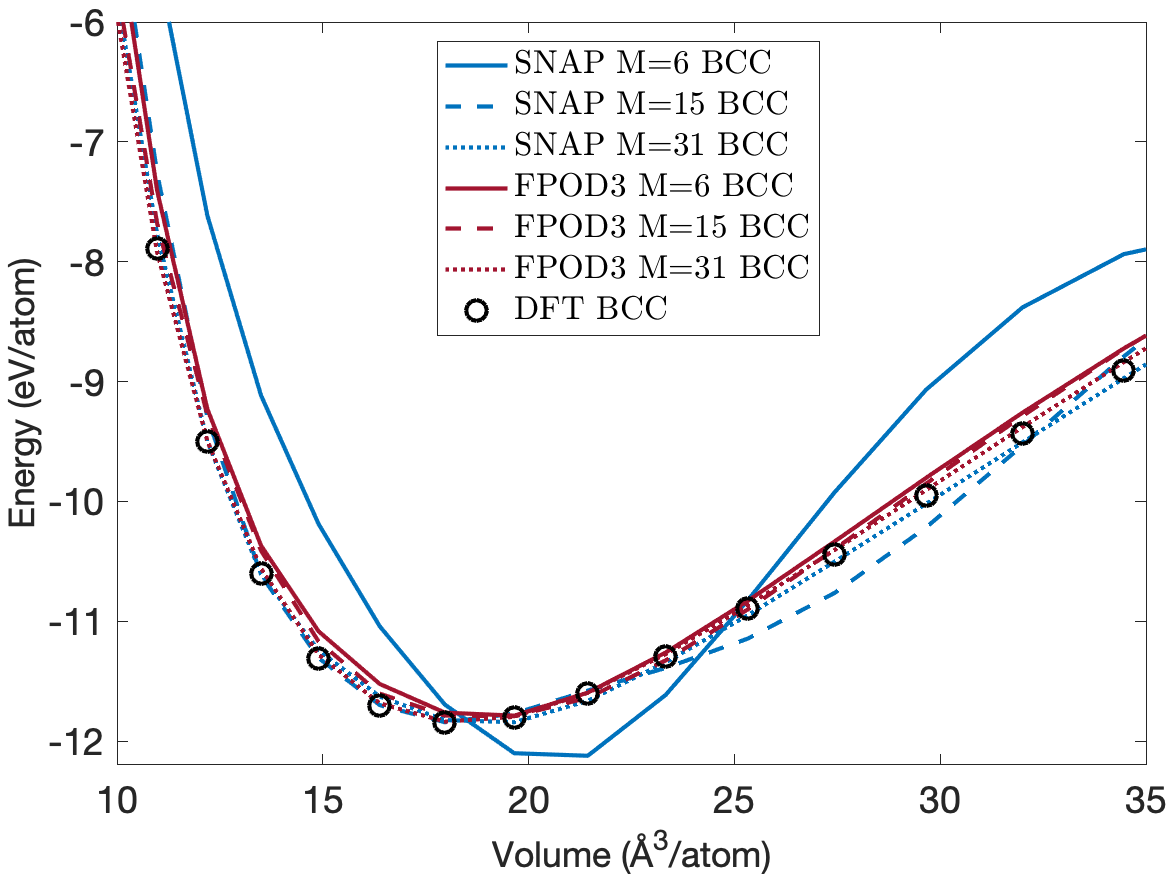}
\caption{\label{fig3a} Energy per atom versus volume per atom for BCC crystal structures for SNAP and FPOD3 potentials in comparison with DFT data.}
\end{figure}

One of the crucial requirements for interatomic potentials is that they predict the correct minimum energy crystal structure and that the energy curve of crystal structures is accurately predicted. Figure \ref{fig3a} plots the energy per atom computed with SNAP and FPOD3 potentials as a function of volume per atom for BCC crystal structures. In addition, energies computed from density functional theory (DFT) are included as cross validation. The BCC phase is the most stable throughout the volume range considered. We see that the energies predicted using SNAP potentials are significantly improved when we increase the number of descriptors. The predicted energy curve of SNAP with $M=31$ is considerably more accurate than that of SNAP with $M=6$. For FPOD3 with just $M=6$ descriptors, the predicted energy curve agrees well with the DFT energy curve of BCC crystal structures. The predicted energy curve of FPOD3 with $M=31$ is much more accurate than that of FPOD3 with $M=6$. However, the improvement of FPOD3 with $M=31$ over $M=6$ is not clearly seen from Figure \ref{fig3a} because of the scale of the plot.    

\begin{figure*}
\includegraphics[scale=0.425]{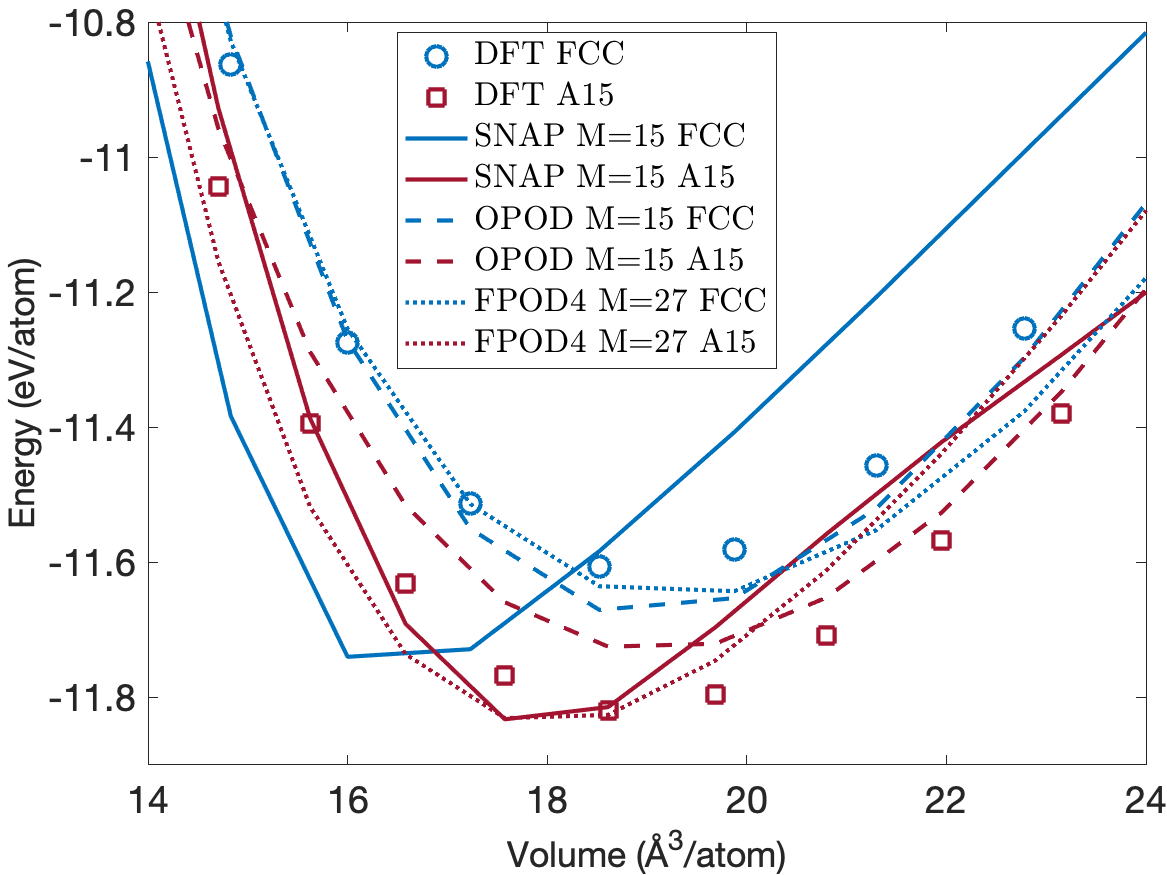} \qquad 
\includegraphics[scale=0.425]{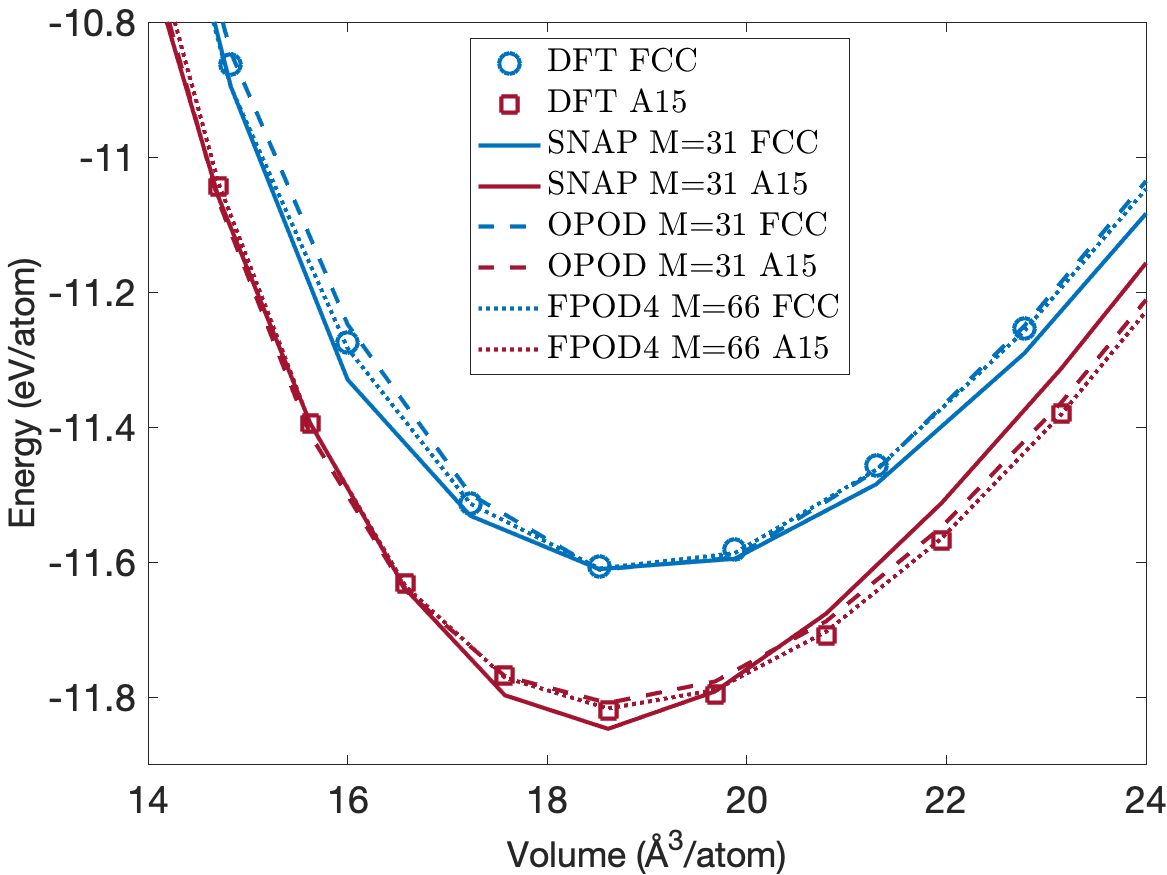}
\caption{\label{fig3} Energy per atom versus volume per atom for FCC and A15 crystal structures for SNAP, OPOD, FPOD4 potentials in comparison with DFT data.}
\end{figure*}

Figure \ref{fig3} depicts the energy per atom computed with SNAP, OPOD and FPOD4 potentials as a function of volume per atom for FCC and A15 crystal structures. In addition, DFT energies are included as cross validation. The FCC phase has a minimum energy about 0.2eV/atom above that of the BCC phase. We again see that the predicted energies are closer to the DFT energies when we increase the number of descriptors for these models. For $M=15$, SNAP predicts the energy curve of A15 phase more accurately than that of FCC phase. For $M=31$, both SNAP and OPOD potentials are in good agreement with DFT calculations for both FCC and A15 phases. FPOD4 tends to yield better predicted energies than SNAP and OPOD. The predicted energy curve of FPOD4 with $M=66$ is almost indistinguishable from the DFT energy curve for both FCC and A15 phases.


\begin{figure}[htbp]
\includegraphics[scale=0.425]{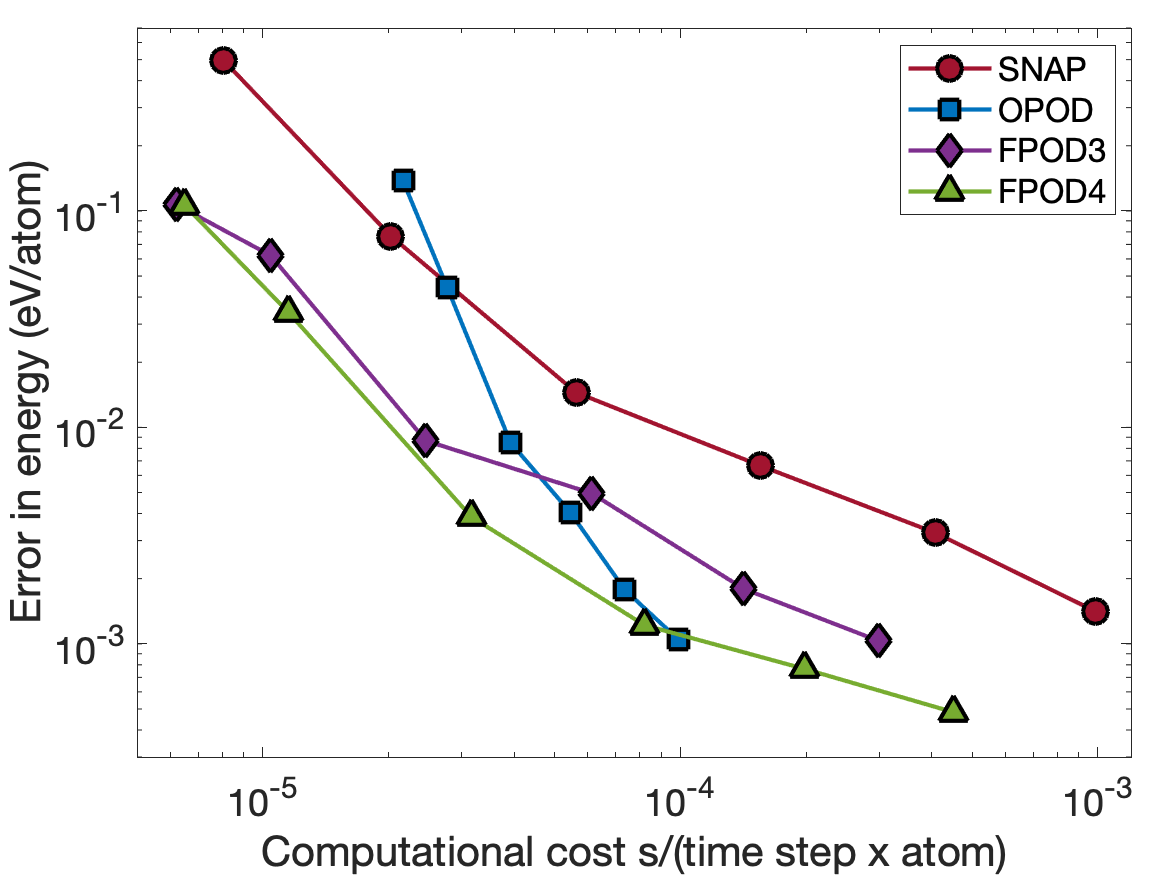}
\caption{\label{fig4} Training error versus the computational cost of MD simulations for the Ta system of 16000 atoms. MD simulations are performed using LAMMPS \cite{Thompson2022} on a single CPU core of Intel i7 2.3 GHz for SNAP, OPOD, FPOD3, FPOD4 with different numbers of descriptors listed in Table \ref{table6}.}
\end{figure}

The number of descriptors has a strong effect not only on the accuracy but also the computational cost of these potentials. The computational complexity of OPOD  is $O(N_i^2 K_r^{(3)}P_a^{(3)})$ per atom, while that of FPOD3 is $O(N_i K_r^{(3)} (P_a^{(3)})^3)$. Note that the computational complexity of FPOD3 scales linearly with $N_i$ and cubically with $P_a^{(3)}$, whereas that of OPOD scales quadratically with $N_i$ and linearly with $P_a^{(3)}$. The computational complexity of FPOD4 is equal to that of FPOD3 plus the cost of computing the four-body descriptors and their derivatives. The computational complexity of SNAP is $O(N_i J^5)$ per atom \cite{Thompson2015}, where $J$ is the upper limit for the index $j$ of $U_{jmm'}(\omega, \phi, \theta)$ in (\ref{4dlap1}). Figure \ref{fig4}
illustrates the trade-off between computational cost and training error for SNAP, OPOD, FPOD3, and FPOD4 potentials. The computational cost is measured in terms of second per time step per atom for molecular dynamics (MD) simulations. These MD simulations are performed using LAMMPS \cite{Thompson2022} on a single CPU core of Intel i7 2.4 GHz with $20 \times 20 \times 20$ bulk supercell containing 16000 Tantalum atoms. It is interesting to compare the computational cost of OPOD and FPOD3 since they yield similar errors for the same number of descriptors. We observe that FPOD3 is faster than OPOD for $M = 6, 15, 31$ and yet slower than OPOD for $M = 56, 92, 141$. We expect that because FPOD3 is faster than OPOD for small values of $P_a^{(3)}$, whereas OPOD is faster than FPOD for large values of $P_a^{(3)}$. FPOD4 outperforms SNAP, OPOD, FPOD3 since it is faster than the other models for the same accuracy. For the same accuracy of $0.01$ eV/atom or less, FPOD4 is more than one order of magnitude faster than SNAP.

\subsection{Discussions}

The above results show that three-body POD potentials tend to yield smaller errors than four-body SNAP potentials for the same number of descriptors. There are two reasons to explain the better accuracy of POD potentials. First, POD potentials include both two-body and three-body descriptors, while SNAP potentials include four-body descriptors. Second, the radial basis functions of POD potentials are constructed from the proper orthogonal decomposition of parametrized potentials, whereas those of SNAP potentials are Gegenbauer polynomials. As a result, POD potentials can capture two-body and there-body interactions better than SNAP potentials. For systems where atomic interactions are prominently two-body and three-body, POD potentials can outperform SNAP potentials. One can improve SNAP potentials by combining two-body and three-body POD descriptors with four-body SNAP descriptors to construct  more accurate potentials. 

OPOD and FPOD3 potentials yield similar errors for the same number of descriptors. They may be viewed as the same model with two different implementations and thus different computational costs. In particular, the cost ratio between OPOD and FPOD3 is $O(N_i/(P_a^{(3)})^2)$. Therefore, we should rely on the cost ratio to decide which model to use for problems at hand. FPOD4 models extend FPOD3 models by including additional four-body descriptors. This extension does not significantly increase the computational cost owing to the atom density representation that allows for efficient implementation of the four-body POD descriptors.  As a result, FPOD4 models outperform all the other models on the Tantalum data set. 

The above potentials are highly flexible and accurate for Ta element. They are flexible because one can generate a collection of different models corresponding to different values of the number of descriptors. Depending on the accuracy required for a particular application, one can choose the most cost-effective model to perform MD simulations. They are highly accurate because they can yield errors below the limits of DFT errors. Despite these impressive features, all the potentials presented herein are still considerably more expensive than a number of established empirical potentials such as the SW potential  and the Tersoff potential. The SW and Tersoff potentials are a lot faster because their functional form can be seen as a sum of one two-body descriptor and one three-body descriptor, whereas SNAP and POD potentials have tens or hundreds of descriptors. Unfortunately, the fixed number of descriptors in these empirical potentials makes themselves inflexible and not accurate enough for many applications. However, one can make use of an empirical potential as a reference potential in the construction of a ML potential in order to reduce the number of descriptors. This idea was proposed in \cite{Thompson2015} and successfully applied to different applications.

In this paper, we consider linear regression to construct SNAP, POD, and FPOD models. Of course, further performance improvement can be achieved by using more sophisticated regression methods such as nonlinear regression and neural networks. While nonlinear models require longer training times than linear models, they often yield more accurate predictions for the same computational cost.  In general, training nonlinear models are still very fast compared to performing MD simulations with these models. On the aspect of training times, we emphasize that it takes only a few seconds to a few minutes to train linear SNAP and POD models on a personal computer. The training time is pretty fast for most models once the QM data set is already available. The most time-consuming part of the training process lies in QM calculations to obtain a high-quality diverse data set. 

\section{Conclusions}
\label{conclusions}

We have introduced fast proper orthogonal descriptors for constructing many-body interatomic potentials. The fast PODs have flavors of both internal coordinate descriptors and atom density descriptors. While the fast PODs are explicitly defined as functions of internal coordinates, they can be recast into atom density descriptors. As a result, the computational complexity of the fast PODs scales linearly with the number of neighbors  irrespective of the body orders. We have introduced the fast PODs of two body, three body, and four body in this paper, yet the approach can be extended to arbitrary body orders. 

We have shown that our approach can be applied to existing empirical potentials to reduce their computational cost. Both the SW potential and the MEAM potential can be evaluated with a cost that scales only linearly with the number of neighbors. However, the approach is not applicable to bond-order potentials such as the Tersoff potential because they do not admit the atom density representation. Most existing empirical potentials are functions of two-body and three-body terms except for the four-body dihedral angle potentials that involve selected groups of atoms rather than a sum over all possible triplets to reduce the computational cost. Our approach  brings about the possibility of constructing new many-body empirical potentials, while maintaining the computational cost that scales linearly with the number of neighbors. For instance, the SW and MEAM potentials can be extended to include four-body terms based on the four-body POD descriptors.

We have discussed the relationship between our method and other internal coordinate methods such as ACSFs and atom PIPs. One can use our method to compute ACSFs and atom PIPs with a computational cost that scales linearly with the number of neighbors, as we have shown how to do that for the original PODs \cite{Nguyen2022}. We proposed to extend the ACSF descriptors with the four-body symmetry function (\ref{eq22d}). This symmetry function yields a completely new set of four-body ACSF descriptors which can improve the accuracy of ACSF neural network potentials, while ensuring linear scaling in terms of the number of neighbors. We discussed the relationship  between our method with atom density methods such as SOAP, SNAP, ACE and MTP. The main advantage of atom density descriptors is that their computational cost scales linearly with the number of neighbors irrespective of the body orders. We have shown that internal coordinate descriptors such as POD, ACSF, and atom PIP can be reformulated into atom density descriptors to have the same advantage. However, there are classes of internal coordinate descriptors that do not admit the atom density representation, and therefore they can be recast into atom density descriptors. In other words, the space of internal coordinate descriptors encompasses the space of atom density descriptors. 

We have discussed commonly used regression methods such as linear regression, nonlinear regression, kernel regression, neural networks, graph neural networks to construct interatomic potentials from invariant descriptors. Each method has its own merits. Linear regression is simple, efficient, and amenable to both uncertainty quantification and active learning. More sophisticated regression methods can yield more accurate models at the expense of adding complexity into training process and evaluation of energies and forces. Neural networks and graph neural networks currently receive significant attention from researchers due to the availability of Pytorch and Tensor Flow frameworks that greatly simplify the implementation of NN and GNN potentials. 

We have developed many-body interatomic potentials by combining the fast POD method with linear regression. The fast POD potentials were demonstrated on DFT data set of Tantalum element and compared with SNAP and POD potentials. All potentials were able to reproduce the DFT reference data set with 1 meV/atom accuracy. POD potentials are significantly faster and more accurate than SNAP potentials. The accuracy and efficiency of the fast POD potentials can be attributed to  the proper orthogonal decomposition and the atom density representation.

\section*{Acknowledgements} 
I would like to thank all members of the CESMIX center at MIT for fruitful and invaluable discussions leading me to the ideas presented in this work. I want to thank Jaime Peraire, Youssef Marzouk, Nicolas Hadjiconstantinou, Dionysios Sema, Yeongsu Cho, William Moses, Jayanth  Mohan, Dallas Foster, Valentin Churavy, Mathew Swisher for the many discussions we have on a wide ranging of different topics related to this work. I would also like to thank Andrew Rohskopf, Axel Kohlmeyer and Aidan Thompson for fruitful discussions about LAMMPS implementation of POD potentials. I  gratefully acknowledge the United States  Department of Energy under contract DE-NA0003965 and the AFOSR Grant No. FA9550-22-1-0356 for supporting this work.

\bibliography{library}

\end{document}